\documentclass[11pt,twoside]{article}

\usepackage[pctex32]{graphicx}
\usepackage{graphicx,epsfig}
\usepackage{epsfig}
\usepackage[figuresright]{rotating}
\usepackage{cite}
\usepackage{latexsym}
\def\tablefootnote#1{%
\hbox to
\textwidth{\hss\vbox{\hsize\captionwidth\footnotesize#1}\hss}}

\usepackage{bm}
\usepackage{latexsym}
\usepackage{dcolumn}
\usepackage{amsmath,amsfonts,amssymb}
\usepackage{psfrag}
\usepackage{amsthm}
\usepackage{wrapfig} 
\usepackage{enumerate} 

\def\be {\begin{equation}}
\def\ee {\end{equation}}
\def\bea {\begin{eqnarray}}
\def\eea {\end{eqnarray}}
\def\bc {\begin{center}}
\def\ec {\end{center}}
\def\bfg {\begin{figure}}
\def\efg {\end{figure}}
\def\bi {\begin{itemize}}
\def\ei {\end{itemize}}
\def\nn {\nonumber}

\def\la {\label}
\def\le {\left}
\def\ri {\right}
\def\pa {\partial}
\def\fr {\frac}
\def\sq {\sqrt}

\def\a  {\alpha}
\def\b  {\beta}
\def\c  {\gamma}

\def\d  {\delta}
\def\D  {\Delta}
\def\e  {\epsilon}
\def\f {\phi}

\def\k  {\kappa}
\def\l  {\lambda}
\def\L  {\Lambda}
\def\m  {\mu}
\def\n  {\nu}

\def\O  {\Omega}
\def\p  {\pi}
\def\r  {\rho}
\def\th {\theta}
\def\s {\sigma}
\def\t  {\tau}
\def\vph {\varphi}

\def\cA {\mathcal A}
\def\cN {\mathcal N}
\def\cH {\mathcal H}
\def\cE {\mathcal E}
\def\xh {\hat{x}}

\def\kt {\tilde{\kappa}}
\def\et {\tilde{\epsilon}}
\newcommand{\eel}[1] {\label{#1}\end{equation}}

\newcommand{\ket}[1]{\mbox{$\mid #1\,\rangle$}}

\newcommand{\SBH}{S_{_{\rm BH}}}
\newcommand{\Sen}{S_{_{\rm ent}}}

\newcommand{\bdm}{\begin{displaymath}}
\newcommand{\edm}{\end{displaymath}}

\newcounter{mylistcount} 
\begin{document}

\title{Black hole entropy from entanglement: A review}

\author{Saurya Das$^1$\footnote{email: saurya.das@uleth.ca},~
S. Shankaranarayanan$^2$\footnote{email: shanki.subramaniam@port.ac.uk} ~ and ~
Sourav Sur$^{1}$\footnote{email: sourav.sur@uleth.ca}\\ \\
{\small \em $^1$~Dept. of Physics, University of Lethbridge,
4401 University Drive, Lethbridge, Alberta, Canada T1K 3M4}\\
{\small \em $^2$~Institute of Cosmology and Gravitation,
University of Portsmouth, Portsmouth P01 2EG, U.K}
}

\date{}
\maketitle

\begin{abstract}
We review aspects of the thermodynamics of black holes and in particular take
into account the fact that the quantum entanglement between the degrees of
freedom of a scalar field, traced inside the event horizon, can be the
origin of black hole entropy. The main reason behind such a plausibility
is that the well-known Bekenstein-Hawking entropy-area proportionality
--- the so-called `area law' of black hole physics --- holds for entanglement
entropy as well, provided the scalar field is in its ground state, or
in other minimum uncertainty states, such as a generic coherent state
or squeezed state. However, when the field is either in an excited state or
in a state which is a superposition of ground and excited states, a power-law
correction to the area law is shown to exist. Such a correction term falls off
with increasing area, so that eventually the area law is recovered for large
enough horizon area. On ascertaining the location of the microscopic degrees of
freedom that lead to the entanglement entropy of black holes, it is found that
although the degrees of freedom close to the horizon contribute most to the
total entropy, the contributions from those that are far from the horizon are
more significant for excited/superposed states than for the ground state. Thus,
the deviations from the area law for excited/superposed states may, in a way,
be attributed to the far-away degrees of freedom. Finally, taking the scalar
field (which is traced over) to be massive, we explore the changes on the
area law due to the mass. Although most of our computations are done in flat
space-time with a hypothetical spherical region, considered to be the analogue
of the horizon, we show that our results hold as well in curved space-times
representing static asymptotically flat spherical black holes with single
horizon.\\

\end{abstract}

\section{Introduction \la{intro}}

One of the most remarkable features of black hole physics is the
realization that black holes behave as thermodynamic systems and
possess entropy and temperature. The pioneering works in the field of
black hole thermodynamics started with Bekenstein \cite{bek},
who argued that the universal applicability of the second law of
thermodynamics rests on the fact that a black hole must possess an entropy
($S_{_{\rm BH}}$) proportional to the area ($\cA_{\rm H}$) of its
horizon. The macroscopic properties of black holes were subsequently
formalized by Bardeen, Carter and Hawking \cite{bch} as the four laws
of black hole mechanics, in analogy with ordinary thermodynamics. They
showed that (i) the surface gravity $\k$, which is the force applied
by an observer at spatial infinity to hold a particle of unit mass in
place at the location of the horizon, is same everywhere on the
horizon for a stationary black hole --- the statement of the zeroth law of black hole
physics. (ii) The surface gravity $\k$ of the black hole analogically
resembles the temperature ($T_{\rm H}$) of the hole, in accordance
with the interpretation of the horizon area as the black hole entropy. This
may be conceived from the first law, which states that the change in
mass (energy) of the black hole is proportional to the surface gravity times
the change in horizon area. 

Hawking's demonstration of black hole thermal radiation \cite{haw} paved the
way to understand the physical significance of the temperature $T_{\rm
H}$ (and hence the entropy-area proportionality). Hawking showed
that quantum effects in the background of a body collapsing to a
Schwarzschild black hole leads to the emission of a thermal radiation at a
characteristic temperature:
\be
\la{H-temp}
T_{\rm H} = \le(\frac{\hbar c}{k_{_{B}}} \ri) \frac{\kappa}{2 \pi} = 
\le(\frac{\hbar c^3}{G k_{_{B}}}\ri) \frac{1}{8 \pi M}\, ,
\ee
where $G$ is the Newton's constant in four dimensions, $k_{_{B}}$ is
the Boltzmann constant, and $M$ is the mass of the black hole.  The factor of
proportionality between temperature and surface gravity (and as such
between entropy and area) gets fixed in Hawking's derivation
\cite{haw}, thus leading to the Bekenstein-Hawking area law:
\be
\la{BH-law}
S_{_{\rm BH}} = \le(\frac{k_{_{B}}}{4}\ri) 
\frac{\cA_{\rm H}}{\ell_{_{\rm Pl}}^2} \quad\, ,
\ee
where ${\ell_{_{\rm Pl}}} = \sqrt{G \hbar/c^3}$ is the four
dimensional Planck length. 

Black-hole thermodynamics and, in particular, black-hole entropy raises 
several important questions which can be broadly classified into two 
categories: 
\begin{itemize}
\item {\bf Gravitational collapse leading to black-hole formation}
\begin{enumerate}[(i)]
\setcounter{enumi}{\value{mylistcount}}
\item  What is the dynamical mechanism that makes $\SBH$ a 
universal function, independent of the black-hole’s past history and 
detailed internal condition? 
\item How does a pure state evolve into a mixed (thermal) state? Is 
there a information loss due to the formation of black-hole and Hawking 
process? Does the usual quantum mechanics need to be modified in 
the context of black-holes? 
\item Can quantum theory of gravity remove the formation of space-time singularity due to 
the gravitational collapse? 
\setcounter{mylistcount}{\value{enumi}}
\end{enumerate}

\item {\bf Near thermodynamical equilibrium}
\begin{enumerate}[(i)]
\setcounter{enumi}{\value{mylistcount}}
\item Unlike other thermodynamical systems, why is black-hole entropy non-extensive? i. e. 
why $\SBH$ is proportional to area and not volume?
\item Why is the black-hole entropy large?\footnote{In order to see that, 
let us compare $\SBH$ with the entropy of the current universe. The entropy of 
the universe within our horizon today \cite{Kolb-Turner} is
\be
S_{_{\rm Univ}} \sim 10^{87} \le(\frac{T h^{-1}}{2.75 K}\ri)
\ee
where $T$ is the temperature of the universe, $h$ is of the order
unity. On the other hand the entropy of a Schwarzschild black-hole is 
\be
\SBH \sim 10^{77} \le(\frac{M}{M_{\odot}}\ri)^2
\ee
where $M_{\odot}$ is the solar mass. Hence, a couple of hundred
thousand solar mass black holes can contain as much entropy as is free
in the entire universe. There is increasing evidence that super-massive
black holes exist at the center of many galaxies. By now we know that
a large fraction of galaxies --- of the total (of order) $10^{11}$ --- 
contain such super-massive black holes with mass range 
$10^6 < M_{_{\rm BH}}/M_{\odot} < 10^{10}$. This implies that 
the entropy of the black holes dominates all other
sources of entropy. Hence, understanding the origins of black-hole 
entropy may help us explain the entropy budget of the universe.}
\item How $S_{_{\rm BH}}$ concords with the standard view of the statistical origin? What 
are the black-hole microstates?
\be
S \stackrel{?}{=} k_{_{B}} \ln \le( {\mbox{\# of microstates}} \ri) \nn
\ee
\item Are there corrections to $\SBH$? If there are, how generic are they? 

\item Where are the microscopic degrees of freedom responsible for black-hole entropy located?
\end{enumerate}

\end{itemize}

These questions often seem related, which a correct theory of quantum
gravity is expected to address. In the absence of a workable theory of
quantum gravity, there have been several approaches which address one
or several of the above questions. Most of the effort in the
literature, as in this review, has been to understand the microscopic
statistical mechanical origin of $\SBH$ assuming that the black-hole
is in a (near) thermal equilibrium or not interacting with surroundings.

The various approaches may broadly be classified into two categories
\cite{entropyrev}: 
(a) the ones that associate $S_{_{\rm BH}}$ with fundamental 
states such as strings, $D$-Branes, spin-networks,
etc. \cite{stringsetc,arom}, and 
(b) the other that associate $S_{_{\rm BH}}$ with quantum fields in a 
fixed BH background, like the brick-wall model \cite{thooft}, the quantum 
entanglement of modes inside and outside of the horizon 
\cite{bkls,sred,sdshankiES,masdshanki} and the Noether charge \cite{Wald:1993a}. 
As mentioned above, although, none of these approaches can be considered 
to be complete; all of them --- within their domains of applicability --- 
by counting certain microscopic states yield (\ref{BH-law}). This is in 
complete contrast to other physical systems, such as ideal gas, where 
quantum degrees of freedom (DOF) are uniquely identified and lead to the 
classical thermodynamic entropy.

The above discussion raises three important questions which we try to address 
in this review: 
\begin{enumerate}
\item {\it Is it sufficient for an approach to reproduce (\ref{BH-law}) or 
need to go beyond $\SBH$?}

As we know, $S_{_{\rm BH}}$ is a semi-classical result and there are
strong indications that Eq. (\ref{BH-law}) is valid for large black
holes [i.e. $\cA_{\rm H} \gg \ell_{_{\rm Pl}}$]. However, it is not
clear, whether this relation will continue to hold for the Planck-size
black-holes. Besides, there is no reason to expect that $\SBH$ to be
the whole answer for a correct theory of quantum gravity. In order to
have a better understanding of black-hole entropy, it is imperative
for any approach to go beyond $\SBH$ and identify the subleading
corrections.
\item  {\it Are the quantum DOF that contribute to $\SBH$ and its subleading 
corrections, identical or different?} 

In general, the quantum DOF can be different. However,
several approaches in the literature \cite{correction} that do lead to
subleading corrections either assume that the quantum DOF are
identical or do not {\it disentangle} DOF contribution to $S_{_{\rm
BH}}$ and the subleading corrections.

\item {\it Can one locate the quantum DOF that give rise to $\SBH$ and 
its subleading corrections?} 

More specifically, can we determine to what extend do the quantum DOF
close to the horizon or far from the horizon contribute to the $\SBH$
and its corrections.  Depending on the approach, one either counts
certain DOF on the horizon, or abstract DOF related to the black hole,
and there does not appear to be a consensus about which DOF are
relevant or about their precise location \cite{entropyrev}.
\end{enumerate}

In this review, using the approach of entanglement of modes across 
the black-hole horizon, we address the above three issues and show that: 
\begin{enumerate}
\item  the entanglement leads to generic power-law corrections to the 
Bekenstein-Hawking entropy, of the form \cite{sdshankiss}:
\be \la{ms-ent}
S = \s_0 \le(\frac{\cA}{a^2}\ri) + \s_1 \le(\frac{\cA}{a^2}\ri)^{- \n} ~;~~~~ 
(0 < \n < 1)\, ,
\ee
where $s_0, \s_1$ are constants and $a$ is the lattice spacing,  
\item the quantum degrees of freedom that lead to $S_{_{\rm BH}}$ and
subleading corrections are different, and 
\item the contribution to $\SBH$ comes from the region close to the horizon while 
the subleading corrections have a larger contribution from the region far from the
horizon \cite{sdshankiDoF}.
\end{enumerate}

This review is organized as follows: in the next section we briefly
review the basic features of quantum entanglement and the concept of
entanglement entropy.  In sec. (\ref{sec:connection}), we provide a 
heuristic picture of the link between the entanglement entropy and 
black-hole entropy. In sec. \ref{ent-sf} we discuss the procedure
and assumptions to compute the entanglement entropy of a scalar field
in black-hole space-times. In sec. \ref{ent-gs} we review the cases
where the scalar field is either in its ground state, or in a
generalized coherent state, or in a class of squeezed states --- for
all these cases the area law is found to hold. In sec. \ref{ent-es}, we
show that the for the superposition of ground and 1-particle state,  
the entanglement entropy has a subleading power-law corrections to the 
AL. In sec. \ref{sf-dof} we study the locations scalar field DOF that are
responsible for the entanglement entropy. In sec. \ref{sf-mass} we
examine the entanglement entropy due to a massive scalar field and
compare it with that obtained for a massless scalar field. We conclude
with a summary and open questions in sec. \ref{concl}. In Appendix
\ref{sf-mot} we discuss the relevance for considering massless or
massive scalar field for computing the entanglement entropy of black holes,
from the perspective of gravitational perturbations in static black hole
space-times. In Appendix \ref{BH-Ham} we discuss the steps to obtain
the Hamiltonian of a scalar field in the static black-hole background, which
in Lema\^itre coordinates, and at a {\it fixed} Lema\^itre time,
reduces to the scalar field Hamiltonian in flat space-time.

Before we proceed, we outline our conventions and notations: We work
in four-dimensions and our signature for the metric is
$(-,+,+,+)$. Hereafter, we use units with $k_{_{B}} = c = \hbar = 1$
and set the Planck mass $M_{_{\rm Pl}}^2 = 1/(16 \pi G)$.  The quantum
field $\varphi$ is a minimally coupled scalar field.

\section{Entanglement entropy  \la{ent}} 

Let us consider a bipartite quantum mechanical system, i.e., a system which can be 
decomposed into two subsystems $u$ and $v$, such that the Hilbert space of the system 
is a tensor product of the subsystem Hilbert spaces:
\be
{\cal H} = {\cal H}_u \otimes {\cal H}_v \, .
\ee
Let $|u_i\rangle$ and $|v_j\rangle$ are eigen-bases which span the Hilbert spaces 
${\cal H}_u$ and ${\cal H}_v$ respectively. Then $|u_i\rangle \otimes |v_j\rangle$ 
forms an eigen-basis in $\cal H$, and in terms of this a generic wave-function 
$|\Psi\rangle$ in ${\cal H}$ can be expanded as:   
\be \la{wf-full}
|\Psi \rangle = \sum_{ij} d_{ij} |u_i \rangle \otimes |v_j\rangle ~\in {\mathcal H} \, .
\ee
The density matrix operator of the whole system defined by: 
\be \la{dm-full}
\rho \equiv |\Psi \rangle \langle \Psi | \, , 
\ee
has the following properties: (i) it is non-negative, i.e., for any vector $|\phi\rangle$,
$\langle\phi|\r|\phi\rangle \geq 0$, (ii) it is self-adjoint ($\r^\dagger = \r$), (iii) 
its trace is unity (Tr$(\r) = 1$), and (iv) it is idempotent ($\r^2 = \r$), which means
that it has only two eigenvalues $p_n = 0, 1$. 

Now the expectation values of the operators which represent the observables in one subsystem, 
say $u$, can be obtained using the {\it reduced} density matrix operator ($\r_u$) for the 
subsystem u, which is the trace of the full density matrix operator ($\r$) over the other
subsystem $v$: 
\be \la{rdm-u}
\r_u = \mbox{Tr}_v (\r) = \sum_l \langle v_l | \r | v_l\rangle 
= \sum_{i,j,k} |u_i \rangle d_{ik} d^\star_{jk} \langle u_j| \, .
\ee

When the correlation coefficients $d_{ij} = 1$ (for all $i,j$), the wave-function 
$|\Psi\rangle$ describing the entire system is just the product of the wave-functions 
$|\Psi_u\rangle$ and $|\Psi_v\rangle$ which describe the subsystems $u$ and $v$ 
respectively, and the full density matrix is $\r = \r_u \r_v$. The reduced density
matrix for the subsystem $u$ (say) is then simply given by $\r_u = |\Psi_u\rangle
\langle\Psi_u$, and bears the same properties as those listed above for the full
density matrix. In general, however, $|\Psi \rangle \neq |\Psi_u\rangle \otimes 
|\Psi_v \rangle$, and the two subsystems are said to be in an {\it entangled} or 
{\it EPR} state\footnote{For details, see the review \cite{timmermans}. }. For such 
entanglement the reduced density matrix ($\r_u$) for the subsystem $u$ is given by 
the general expression (\ref{rdm-u}). A similar expression can be obtained for the 
reduced density matrix ($\r_v$) for the subsystem $v$, by taking the trace of full 
density matrix ($\r$) over the subsystem $u$. Both $\r_u$ and $\r_v$ have the same 
properties as listed above for $\r$, except that the idempotency is lost, i.e., 
$\r_u^2 \neq \r_u, \r_v^2 \neq \r_v$. As such the eigenvalues of $\r_u$ and $\r_v$ 
are no longer restricted to $0$ and $1$, but are in between $0$ and $1$. However, an 
important property of reduced density matrices is that irrespective of the tracing 
over the eigenvalues remain the same, i.e., both $\r_u$ and $\r_v$ have the same set 
of eigenvalues. 

The {\it entanglement entropy} or {\it Von Neumann entropy} is a manifestation of 
one's ignorance resulting from tracing over one part of the system, and is defined by  
\be \la{ee}
S_{\rm ent} \equiv - Tr_u\le( \r_u \ln \r_u \ri) \equiv - Tr_v\le( \r_v \ln \r_v \ri) 
= - \sum_n p_n \ln p_n ~;~~~~~~ (0 < p_n < 1) \, .
\ee
Since the eigenvalues $p_n$ of $\r_u$ and $\r_v$ are the same, $S_{\rm ent}$ is the 
same regardless of which part of the system is being traced over. 

\section{Entanglement entropy and black-hole entropy --- Connection}
\label{sec:connection}
\begin{wrapfigure}{l}{0.5\textwidth}
  \vspace{-20pt}
  \begin{center}
    \includegraphics[width=0.48\textwidth,height=91mm]{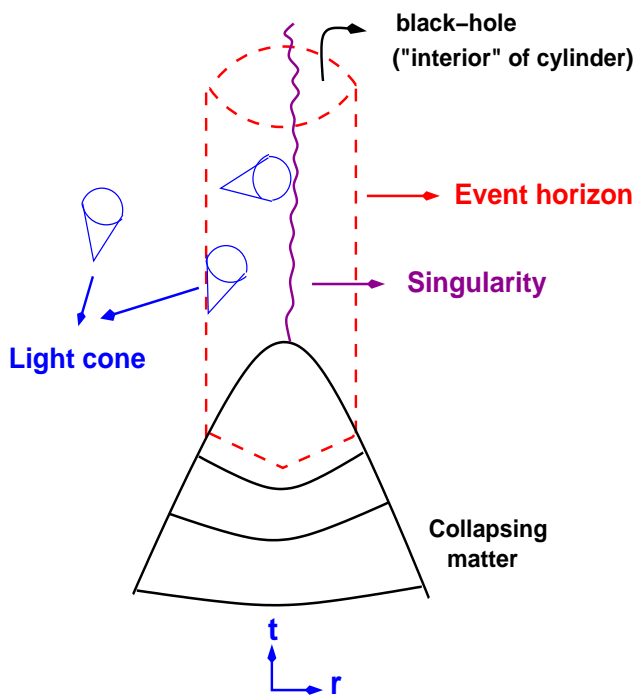}
  \end{center}
  \vspace{-10pt}
  \caption{Space-time diagram depicting the collapse of a star to 
form a black-hole.}
  \vspace{-10pt}
\end{wrapfigure}

Before we proceed with the setup and assumption, in this section, we provide a
heuristic link between the entanglement entropy and black-hole entropy. 

To understand this, let us consider a scalar field on a background of a collapsing 
star. Before the collapse, an outside observer, atleast theoretically, has
all the information about the collapsing star. Hence, the entanglement entropy is 
zero. During the collapse and once the horizon forms, $\SBH$ is non-zero. 
The outside observers at spatial infinity do not have the information about the 
quantum degrees of freedom inside the horizon. Thus, the entanglement entropy is 
non-zero. In other words, both the entropies are associated with the existence 
of horizon\footnote{It is important to understand that it is possible to 
obtain a non-vanishing entanglement entropy for a scalar field in flat space-time 
by artificially creating a horizon \cite{sred}. However, in the case of black-hole,
the event horizon is a physical boundary beyond which the observers do not have access 
to information.}. 

Beside this, $\Sen$ and $\SBH$ are pure quantum effects with no classical analogue. 
Based on this picture, we obtain $\Sen$ of scalar fields in a fixed background.
Although, this analysis is semiclassical, since the entanglement is a quantum effect 
and should be present in any theory of quantum gravity, the results presented here 
do have implications beyond the semiclassical regime. 

\section{Entanglement entropy of scalar fields --- Assumptions and setup}        
\la{ent-sf}

In this section, we consider a free scalar field propagating in spherically symmetric black-hole 
space-times. The motivation for such scalar field is presented in Appendix \ref{sf-mot}] 

We consider a massless scalar field $(\varphi)$ propagating in an asymptotically flat, 
four-dimensional black-hole background given by the Lema\^itre line-element \ref{bh-metric}. 
The Hamiltonian of the scalar field propagating in the above line-element is
\begin{equation}
H_{_{\rm BH}}(\tau) = \sum_{lm} \frac{1}{2} \int_{\tau}^{\infty} \!\!\!\!
d\xi \le[ \frac{1}{r^2 \sqrt{1 - f(r)}} \Pi_{_{lm}}^2
+ \frac{r^{2}}{\sqrt{1 - f(r)}} \le(\pa_{\xi} \vph_{_{lm}}\ri)^2
 + l(l + 1)\sqrt{1 - f(r)} \, \vph_{_{lm}}^2 \ri] \,  ,
\label{eq:ham1}
\end{equation}
where $\varphi_{lm}$ is the spherical decomposed field and $\Pi_{lm}$
is the canonical conjugate of $\varphi_{lm}$ i. e.,
\be
\varphi_{lm} (r) = r \int d\O~Z_{lm} (\theta,\phi) \varphi (\vec r) ~;~~~
\Pi_{lm} (r) = r \int d\O~Z_{lm} (\theta,\phi) \Pi (\vec r) \, , 
\ee
where $Z_{lm} (\theta,\phi)$ are the real spherical harmonics. [For 
discussion about the Lema\^itre coordinates and explicit derivation 
of Eq. (\ref{eq:ham1}), see Appendix (\ref{BH-Ham}).] Note that the 
above Hamiltonian is explicitly time-dependent.

Having obtained the Hamiltonian, the next step is quantization. We use 
Schr\"odinger representation since it provides a simple and 
intuitive description of vacuum states for time-dependent Hamiltonian.
Formally, we take the basis vector of the state vector space to be the eigenstate 
of the field operator ${\hat \varphi}(\tau, \xi)$ on a fixed $\tau$ 
hypersurface, with eigenvalues $\varphi(\xi)$ i. e.
\be
\label{FSE-states}
{\hat \varphi}(\tau,\xi)\ket{\varphi(\xi),\tau} = \varphi(\xi)\ket{\varphi(\xi),\tau}
\ee 
The quantum states are explicit functions of time and are represented
by wave functionals $\Psi[\varphi(\xi),\tau]$ which satisfy the 
functional Schr\"odinger equation:
\be
\label{schrod}
i \frac{\partial \Psi}{\partial \tau} = \int_{\tau}^{\infty} \!\!\!
d\xi  \, H_{_{\rm BH}}(\tau) \, \Psi[\varphi(\xi),\tau] \, .
\ee

We, now, assume that the above Hamiltonian evolves adiabatically. Technically, 
this implies that the evolution of the late-time modes leading to Hawking particles 
are negligible. In the Schroedinger formulation, the above assumption translates to 
$\Psi[\varphi(\xi), \tau]$ being independent of time. At a fixed Lema\^itre time,
Hamiltonian (\ref{eq:ham1}) reduces to the following flat space-time Hamiltonian 
[see Appendix (\ref{BH-Ham}) for details]
\be \la{sf-ham2}
H = \sum_{lm} H_{lm} = \sum_{lm} \fr{1}{2} \int_0^\infty dr \le\{ \p_{lm}^2(r) 
+ x^2 \le[ \fr{\pa}{\pa r} \le( \fr{\varphi_{lm} (r)}{r}\ri) \ri]^2 
+ \fr{l(l+1)}{r^2}~\varphi_{lm}^2(r) \ri\} \, . 
\ee
The quantum states [defined above in Eq. (\ref{FSE-states})] of this Hamiltonian is 
time-independent and $\Psi[\varphi]$ satisfies the time independent Scr\"odinger equation 
\be
\int_{0}^{\infty} dr  H \Psi[\varphi(r)] = E \Psi
\ee

The procedure of finding the entropy involves the following steps: \\ \\
(i) {\it Discretize Hamiltonian (\ref{sf-ham2})}:  
Obtaining an analytic expression for the entropy using the Von Neumann definition (\ref{ee}) 
is prohibitively difficult. For the field theory, even if we obtain closed-form expression 
of the density matrix, it is not possible to analytically evaluate the entanglement entropy. 
Hence, we discretize the Hamiltonian in a spherical lattice of spacing $a$ such that 
$r \rightarrow r_i;~ r_{i+1} - r_i = a$. The ultraviolet cutoff is therefore $M = a^{-1}$.. 
The lattice is of very large but finite size $L = 
(N+1)a ~(N \gg 1)$, with a chosen closed spherical region of radius $R (n + 1/2)a$ 
inside of it, as shown in the Fig. \ref{fig:ee-setup}. It is this closed region, by 
tracing over the inside or outside of which one can obtain the reduced density matrix.
We demand that the field variables $\varphi_{lm} (r) = 0$ for $r \geq L$ so that the
infrared cutoff is ${\tilde M} = L^{-1}$.The ultraviolet
cutoff is therefore $M = a^{-1}$.

Now, in discretizing the terms containing the derivatives in the Hamiltonian (\ref{sf-ham2}), 
one usually adopts the middle-point prescription, i.e., the derivative of the form $f(x) 
d_x[g(x)]$ is replaced by $f_{j + 1/2} [g_{j + 1} - g_j]/a$. The discretized Hamiltonian
is given by
\bea \label{disc1}
H_{lm} = \fr 1 {2a} \sum_{j=1}^N \le[ \p_{lm,j}^2 + \le(j + \fr 1 2\ri)^2 
\le(\fr{\varphi_{lm,j}}{j} - \fr{\varphi_{lm,j+1}}{j+1}\ri)^2 + \fr{l(l+1)}{j^2}~
\varphi_{lm,j}^2 \ri]  \quad , \quad H = \sum_{lm} H_{lm}
\eea
%
\begin{wrapfigure}{l}{0.43\textwidth}
  \vspace{-20pt}
  \begin{center}
    \includegraphics[width=0.40\textwidth,height=57mm]{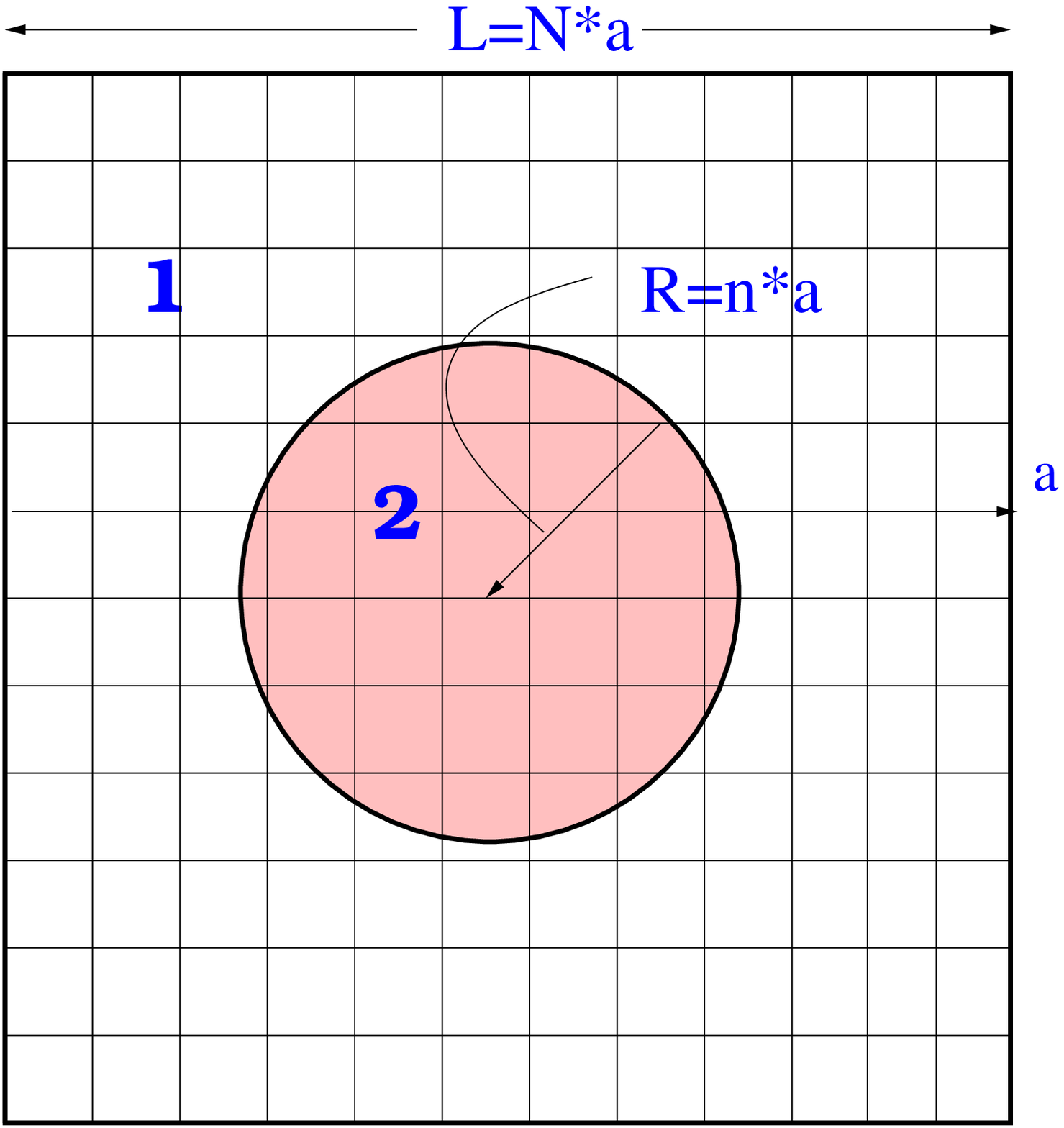}
  \end{center}
  \vspace{-10pt}
  \caption{Discretization of the scalar field propagating 
in the flat space-time. The tracing is done over the shaded region.}
  \vspace{-1pt}
\la{fig:ee-setup} 
\end{wrapfigure}
where $\varphi_{lm,j} \equiv \varphi_{lm}(r_j), ~ \p_{lm} \equiv \p_{lm,j}(r_j)$, which
satisfy the canonical commutation relations:
\be \la{comm-rel}
[\varphi_{lm,j}, \p_{l'm',j'}] = i \d_{l l'}\d_{m m'}\d_{j j'} \, . 
\ee
Up to an overall factor of $a^{-1}$, the Hamiltonian $H_{lm}$, given by Eq.~(\ref{disc1}),
represents the Hamiltonian of $N-$coupled harmonic oscillators (HOs):
\be \la{coupledham1}
H_{_{(N- \rm HO)}} ~=~ \fr 1 2 \sum_{i=1}^N p_i^2 ~+~ \fr 1 2 \sum_{i,j=1}^N x_i K_{ij} x_j \, ,
\ee
where the coordinates $x_i$ replace the field variables $\varphi_{lm}$, the momenta $p_i$
replace the conjugate momentum variables $\pi_{lm}$, and the $N \times N$ matrix $K_{ij}$ 
($i,j = 1,\dots,N$) represents the potential energy and interaction between the oscillators: 
\bea \la{kij}
K_{ij} &=&  \fr 1 {i^2} \le[l(l+1)~\delta_{ij} + \fr 9 4 ~\d_{i1} 
\d_{j1} + \le( N - \fr 1 2\ri)^2 \d_{iN} \d_{jN} + \le\{ \le(i + \fr 1 2\ri)^2 + 
\le(i - \fr 1 2\ri)^2 \ri\} \d_{i,j(i\neq 1,N)}\ri] \nn\\ 
&&- \le[\fr{(j + \fr 1 2)^2}{j(j+1)}\ri] \delta_{i,j+1} - 
\le[\fr{(i + \fr 1 2)^2}{i(i+1)}\ri] \delta_{i,j-1} .
\eea
The last two terms which originate from the derivative term in Eq. (\ref{sf-ham2}) 
denote the nearest-neighbor interactions. Schematically, the matrix $K$ is as shown
below, where the off-diagonal terms represent interactions:
{\small
\bea
K  = \le( \begin{array}{lllllll} 
{\times} & {\times} & {}& {}& {} & {} & {} \\
{\times} & {\times} & {\times} & {}& {} & {} & {} \\
{}  & {\times} & {\times} & {\times} & {} & {} & {} \\
{} & {}  & {\times} & {\times} & {\times} & {} & {} \\
{} & {} & {}  & {\times} & {\times} & {\times} & {}  \\
{} & {} & {} & {}  & {\times} & {\times} & {\times}  \\
{} & {} & {} & {} & {}  & {\times} & {\times}  
\la{mat1} \\
\end{array} \ri) \, .
\eea
}

\noindent
(ii) {\it Choose the quantum state of the field}: 
The most general eigen-state of the Hamiltonian (\ref{coupledham1}) for the $N-$coupled 
HOs is given by
\be \la{excwavefn1}
\psi (x_1,\dots,x_N) = \prod_{i=1}^N \cN_i ~\cH_{\n_i} 
\le(k_{Di}^{1/4}~{\underbar x}_i \ri)
\exp\le( -\fr 1 2 k_{Di}^{1/2}~{\underbar x}_i^2 \ri), 
\ee
where $\cN_i$'s are the normalization constants given by
\be \la{normconst}
\cN_i = \fr{k_{Di}^{1/4}}{\p^{1/4}~\sqrt{2^{\n_i} \n_i!}} \quad, 
\quad (i = 1, \dots N) \, ,
\ee
${\underbar x} = Ux$, ($U^TU=I_N$), $x^T = (x_1,\dots,x_N)$, 
${\underbar x}^T = ({\underbar x}_1,\dots,{\underbar x}_N)$,
$K_D \equiv U K U^T$ is a diagonal matrix with elements $k_{Di}$, 
and $\nu_i \, (i=1 \dots N)$ are the indices of the Hermite 
polynomials ($\cH_{\nu}$). The frequencies are ordered such that 
$k_{Di} > k_{Dj}$ for $i > j$. 

The difficulty however is that it is not possible to work with an arbitrary 
$N-$particle state as the density matrix (\ref{denmatgen1}) cannot be expressed 
in a closed form. Therefore one has to make some specific choices for the quantum
state in order to make the calculations tractable. In the following two sections 
we will discuss several possible choices.

\noindent
(iii) {\it Obtain the reduced density matrix by tracing over certain closed region}: 
The reduced density matrix is obtained by tracing over the first $n$ of the $N$ 
oscillators:
\bea \la{denmatgen1}
\r \le(t; t'\ri) &=& \int \prod_{i=1}^n dx_i ~ 
\psi(x_1,\dots,x_n; t) ~\psi^\star(x_1,\dots,x_n; t') \nn \\
&=& \int \prod_{i=1}^n dx_i \exp \le[-\fr{x^T \O x} 2\ri] 
\prod_{i=1}^{N} \cN_i \cH_{\n_i} \le(k_{Di}^{1/4} 
{\underbar x}_i\ri) \exp \le[-\fr{x'^T \O x'} 2\ri] \prod_{j=1}^{N} 
\cN_j \cH_{\n_j} \le(k_{Di}^{1/4} {\underbar x}'_i\ri)
\eea
where $\O = U^T K_D^{1/2} U$ is an $N \times N$ matrix, such that 
$|\O| = |K_D|^{1/2}$, and we have made the following change in notation: 
$x^T = (x_1, \dots ,x_n ; t_1, \dots, t_{N-1}) = (x_1, \dots, x_n ; t)$, with 
$t \equiv t_1, \dots, t_{N-n} ~; t_j \equiv x_{n+j}, ~j =1, \dots, (N-n)$. 
One may verify that $\r^2 \neq \r$, i.e., the state obtained by integrating over 
$n$ of the HOs is mixed, although the full state is pure.\\

\noindent
(iv) {\it Compute the entropy:} The entanglement entropy can be 
calculated by substituting the reduced density matrix (\ref{denmatgen1}) into
the expression (\ref{ee}).


\section{Warm up --- Entanglement entropy for (displaced) ground state \la{ent-gs}}

In this section, we review the procedure for obtaining $\Sen$ in the ground state, 
coherent state (which is a displaced ground state) and a class of squeezed 
state (which are unitarily related to the ground state).  We show that the 
in these three cases that the entanglement entropy is proportional to area. 

\bigskip
\noindent
\underline{{\large \it Ground State (GS)}:}  

\bigskip

When all the HOs are in their GS, then by setting $\n_i = 0,~$ for all $i$, the wave 
function (\ref{excwavefn1}) takes the form 
\bea \la{gs-wavefn}
\psi_{_{\rm GS}} (x_1, \dots, x_N) ~=~ \prod_{i=1}^N  \le(\fr{k_{Di}}{\p}\ri)^{1/4} 
\exp \le(- \fr 1 2 k_{Di}^{1/2} {\underbar x}_i^2\ri) ~=~ \le(\fr{|\O|}{\p^N}\ri)^{1/4} 
\exp \le[-~ \fr{x^T \cdot \O \cdot x} 2\ri] \, .
\eea
The corresponding density matrix can be evaluated exactly as \cite{sred}:
\be \la{gs-den}
\r_{_{\rm GS}} (t; t') = \sq{\fr{|\O|}{\pi^{N-n} |A|}} ~\exp \le[- \fr{t^T \c t 
+ t'^T \c t'} 2 ~+~ t^T \b t'\ri] \, ,
\ee
where we have decomposed
\bea \la{Omega}
\O  \sim K^{1/2} ~=~ \le( \begin{array}{ll} 
{A} & {B} \\
{B^T} & {C} 
\end{array} \ri) \, ,
\eea
and defined
\be
\b ~=~ \frac{B^T A^{-1} B} 2 ~;~~~ \c ~=~ C - \b  \, .
\ee
$A$ is an $n \times n$ symmetric matrix, $B$ is an $n \times
(N-n)$ matrix, and $C, \b, \c$ are all $(N-n) \times (N-n)$ symmetric
matrices. The matrices $B$ and $\b$ are non-zero only when
the HOs are interacting. 

Performing a series of unitary transformations:
\bea \la{diag}
&& V \c V^T = \c_D = \mbox{diag}~,~ {\bar\b} \equiv \c_D^{- 1/2} 
V \b V^T \c_D^{- 1/2}~,\nn\\ 
&& W {\bar\b} W^T = {\bar\b}_D = \mbox{diag}~,~  v \equiv 
W^T \c_D^{1/2} V ~, 
\eea
one can reduce $\r_{_{\rm GS}} (t; t')$ to a product of the reduced density
matrices $\r_{_{(2-\rm HO)}} (t; t')$ for $(N-n)$ two coupled HOs with 
one oscillator traced over (i.e., $N=2, n=1$) \cite{sred}: 
\be \la{gs-den1}
\r_{_{\rm GS}} (t; t') = \prod_{i=1}^{N-n} \r_{_{(2-\rm HO)}} (t; t')
\quad, \quad  \r_{_{(2-\rm HO)}} (t; t') = \sq{\fr{|\O|}{\pi^{N-n} |A|}} 
\exp \le[- \fr{v_i^2+v_i'^2}{2} + {\bar\b}_i v_i v_i' \ri] \, ,
\ee
where $v_i \in v$ and ${\bar\b}_i \in \bar\b$. Correspondingly, the total 
entanglement entropy is a sum of $(N-n)$ two-HO entropies $S_i^{(2-\rm HO)} ,
~ (i = 1, \dots, N-n)$ which are obtained using the Von Neumann relation
(\ref{ee}) as \cite{sred}:
\be \la{2HO-ent}
S_i^{(2-\rm HO)} = - \ln[1-\xi_i] - \fr{\xi_i}{1 - \xi_i} \ln\xi_i \quad, 
\quad  \xi_i = \fr{{\bar \b}_i}{1+ \sqrt{1 - {\bar \b}_i^2}} \, .
\ee

The total GS entropy for the full Hamiltonian $H = \sum_{lm} H_{lm}$, 
Eq.(\ref{disc1}), is therefore given by
\be \la{gs-ent}
S_{_{\rm GS}} (n,N) = \sum_{lm} S_{lm} (n,N) = \sum_{l=0}^{l_{max}} (2 l + 1)
S_l (n,N)  
\quad, \quad S_l (n,N) = - \ln[1-\xi_l] - \fr{\xi_l}{1 - \xi_l} \ln\xi_l \,
\ee
where $(2 l + 1)$ is the degeneracy factor found by summing over all values of $m$
for the Hamiltonian. Ideally the upper bound $l_{max}$ of the sum should be infinity, 
however in practice one assigns a very large value of $l_{max}$ to compute the entropy 
upto a certain precision. The precision limit $Pr$ is set by demanding $l_{max}$ to be 
such that the percentage change in entropy for a change in $l$ by a step size $l_{st}$, 
never exceeds $Pr$, i.e.,
\be \la{pr-goal}
\vline\fr{S(l_{max}) - S(l_{max} - l_{st})} {S(l_{max} - l_{st})}\vline 
\times 100 ~<~ Pr \, .
\ee
Thus $Pr = 0.01$ (say) implies that the numerical error in the computation of the total 
entropy is less than $0.01\%$.   
%
\begin{wrapfigure}{r}{0.5\textwidth}
  \vspace{-10pt}
  \begin{center}
    \includegraphics[width=0.48\textwidth,height=80mm]{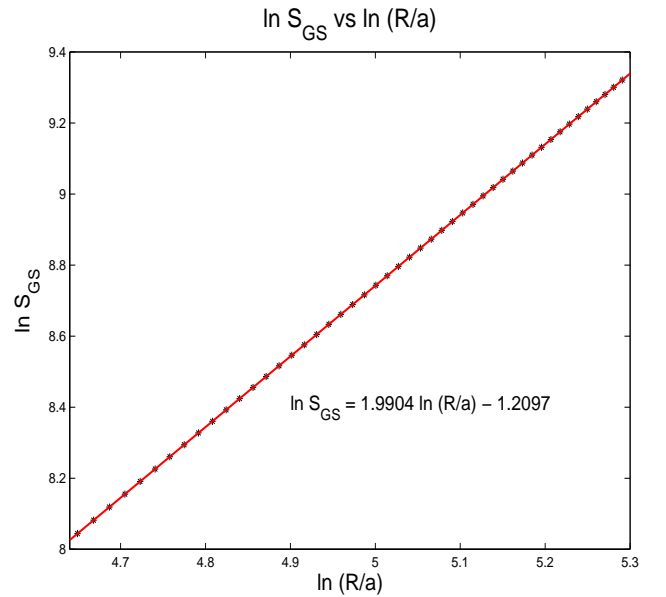}
  \end{center}
  \vspace{-20pt}
  \caption{\small Plot of logarithm of the GS entropy versus $\ln(R/a)$, where $R = (n + 1/2) a$ 
radius of the hypothetical sphere (horizon), for $N = 300, ~ n = 100 - 200$. The graph
is a straight line with slope $\simeq 2$.}
  \vspace{-1pt}
\la{fig:GS}
\end{wrapfigure}
It has been shown in ref. \cite{sred} that for $N \simeq 60$ and $n \leq N/2$, the
GS entropy computed numerically follows the relation:
\be \la{sredresult}
S_{_{\rm GS}} = 0.3 \le(n + \fr 1 2\ri)^2 = 0.3 \le(\fr{R}{a}\ri)^2 = 
\fr {0.3}{4 \pi} \le(\fr {\cA}{a^2}\ri) \, ,
\ee
where $R = (n + 1/2) a$ is the radius of the hypothetical spherical surface (which is an 
analogue of the horizon in flat space-time) the DOF inside or outside of which are traced 
over, and $\cA = 4 \pi R^2$ is the surface area of this sphere. For a fixed value of $n$,
if $N$ is varied the GS entropy remains the same. This implies that $S_{_{\rm GS}}$ does 
not depend on the infrared cutoff ${\tilde M} = L^{-1}$, where $L = (N + 1) a$, although 
it depends on the ultraviolet cutoff $M = a^{-1}$:~ $S_{_{\rm GS}} = 0.3 M^2 R^2$ \cite{sred}. 

In Fig. \ref{fig:GS} we have plotted $\ln S_{_{\rm GS}}$ vs $R/a = n + 1/2$ for the
parametric values: $N = 300$ and $n = 100 - 200$. The data fits into a straight line
with slope very close to $2$ (for a precision limit $Pr = 0.1 \%$ in the computation 
performed using MATLAB has been). The area law ($S_{_{\rm GS}} \propto \cA$) is thus
found to hold generically for GS. 

\bigskip
\noindent
\underline{{\large \it Generalized Coherent State (GCS)}:}  

\bigskip

We now examine the situation where all the oscillators are in GCS, instead of being
in GS. Similar to GS, a GCS is a minimum uncertainty state ($\D x \D p = 1/2$). 
However unlike GS, the GCS is not an energy eigenstate of HO, it is an eigenstate of
the HO annihilation operator with real eigenvalue. The GCS wave-function 
$\psi_{_{\rm GCS}}$ differs from the GS wave-function, Eq. (\ref{gs-wavefn}), by 
constant {\it shifts} $~\a_i$ in the coordinates $\underbar x_i$, as:
\be \la{gcs-wavefn}
\psi_{_{\rm GCS}} (x_1, \dots, x_N) ~=~ \prod_{i=1}^N  \le(\fr{k_{Di}}{\p}\ri)^{1/4} 
\exp \le[- \fr 1 2 k_{Di}^{1/2} \le({\underbar x}_i - \a_i\ri)^2\ri] ~\sim~ 
\prod_{i=1}^N \exp \le(- i {\underbar p}_i \a_i\ri) \exp \le(- \fr 1 2 k_{Di}^{1/2}
{\underbar x}_i^2\ri] \, .
\ee  
where ${\underbar p}_i = - i \pa/\pa {\underbar x}_i$ are the momenta. Physically, the 
real and imaginary parts of the $N$ complex GCS parameters $\a_i$ correspond to the 
classical position ($x_0$) and momentum ($p_0$) of the individual HOs respectively, i.e., 
$\a_i = x_0 - i p_0/k_{Di}$. Defining shifted coordinate variables \cite{sdshankiES}:
\be \la{gcs-x}
{\tilde x } \equiv  x - U^{-1}\a \quad, \quad d\tilde x = dx \, ,
\ee
one can show that 
\be
\psi_{_{\rm GCS}} (x_1, \dots, x_N) = \le(\fr{|\O|}{\p^N}\ri)^{1/4} 
\exp \le[- \fr{{\tilde x}^T \cdot \O \cdot {\tilde x}}{2} \ri]
= \psi_{_{\rm GS}} ({\tilde x}_1, \dots, {\tilde x}_N) \, . 
\ee
Thus the GCS wave-function is of the same form as the GS wave-function, albeit in 
terms of the shifted variables. As such, the reduced density matrix $\r_{_{\rm GCS}}$
for GCS will also be of the same form as that for GS, Eq.(\ref{gs-den}), with the
variables $(t;t')$ replaced by $({\tilde t};{\tilde t}')$, where ${\tilde t} \equiv 
{\tilde t}_1, \dots, {\tilde t}_{N-n} ~; {\tilde t}_j \equiv {\tilde x}_{n+j}, ~j =
1, \dots, (N-n)$:
\bea \la{gcs-den}
\r_{_{\rm GCS}} \le(t; t'\ri) = \int \prod_{i=1}^n dx_i ~ \psi_{_{\rm GCS}}
(x_1,\dots,x_n; t) ~\psi_{_{\rm GCS}}^\star(x_1,\dots,x_n; t') 
= \r_{_{\rm GS}} \le({\tilde t}; {\tilde t}'\ri) \, .
\eea
Consequently, the eigenvalues of the density matrix, and hence the entropy for GCS
will be the same as those for GS \cite{sdshankiES}. Thus the area law holds for the 
case of GCS as well.

\bigskip
\noindent
\underline{{\large \it Squeezed State (SS)}:}  

\bigskip

Let us now consider the case where all the HOs are in a class of SS. The squeezed 
states are also minimum uncertainty packets ($\D x \D p = 1/2$), similar to GS and 
GCS. However, there is a squeezing either in the positions ($\D x \ll 1, \D p \gg 1$)
or in the momenta ($\D p \ll 1, \D x \gg 1$). We consider a class of SS, which is
characterized by a unique squeezing parameter $\zeta$, and described by the
wave-function
\be \la{ss-wavefn}
\psi_{_{\rm SS}} (x_1, \dots, x_N) ~\sim~ \zeta^{N/2} \prod_{i=1}^N  \exp \le[- 
\zeta \sum_{i=1}^N \fr 1 2 k_{Di}^{1/2} {\underbar x}_i^2\ri] \, .
\ee  
In this case, defining scaled coordinate variables \cite{sdshankiES}:
\be \la{ss-x}
{\tilde x } \equiv  \sqrt{\zeta}~ {\underbar x} \quad, \quad d\tilde x = \sqrt{\zeta}
~d{\underbar x} \, ,
\ee
one finds that the SS wave-function and the corresponding density matrix reduce to
the forms same as those for GS, albeit with the replacement $x \rightarrow {\tilde 
x}$. As such, upto an irrelevant multiplicative factor, the SS entropy turns out to
be the same as the GS entropy, Eq. (\ref{sredresult}), and hence the area law holds.
   
Thus, in all the above cases where we have the states which form minimum uncertainty 
packets, the entanglement entropy is proportional to the area. In the next section,
we show that if we consider a higher excited state, the area law is not robust. 
In particular, we show that the superposition of the ground and excited state leads
to power-law corrections to area.

\section{Power-law corrections to the area-law}
\la{ent-es}

Ideally the most interesting thing would be to find the entanglement entropy for
a general eigenstate, given by Eq. (\ref{excwavefn1}), which may be the ground
state or the first, second, ..., etc. excited states, or a superposition of such 
states. However, as mentioned earlier, the density matrix cannot be written 
in a closed form for such a general state. We therefore resort to a simple class 
of excited states, viz., (i) 1-particle excited state and (ii) superposition 
of GS and 1-particle excited state. In the case of 1-particle excited state, 
we show that the entanglement entropy {\it does not} scale as area while in 
the superposed state the entanglement entropy has power-law corrections to the 
area-law.

\subsection{1-Particle Excited state}
\la{ent-es1}

1-particle excited state is described by a wave-function 
$\psi_{_{\rm ES}}$ which is a linear superposition of $N$ HO wave functions, each 
of which has {\it exactly one} HO in the first ES and the rest $(N-1)$ in their 
GS \cite{masdshanki,sdshankiES}. Using Eq. (\ref{excwavefn1}), such an 1-particle
ES wave-function is expressed as:
\bea \la{es-wavefn}
\psi_{_{\rm ES}} (x_1, \dots, x_N) &=& \sum_{i=1}^N \le(\fr{k_{Di}}{4 \p}\ri)^{1/4} 
\a_i \cH_1 \le(k_{Di}^{1/4} {\underbar x}_i\ri) \exp \le(-\fr 1 2 \sum_{j} 
k_{D j}^{1/2}~{\underbar x}_j^2\ri) \nn\\
&=& \sqrt{2} \le(\a^T K_D^{1/2} {\underbar x}\ri) \psi_{_{\rm GS}} \le(x_1, \dots, 
x_N\ri),
\eea
where $\a^T = \le(\a_1, \dots, \a_N\ri)$ are the expansion coefficients, and 
$\a^T \a = 1$ so that $\psi_{_{\rm ES}}$ is normalized.

From Eq.(\ref{denmatgen1}) one finds the reduced density matrix $\r_{_{\rm ES}}
(t; t')$ for ES as:
\bea \label{es-den}
\r_{_{\rm ES}} (t; t') ~=~ \int \prod_{i=1}^n dx_i \psi_{_{\rm ES}} \le(x_i; t\ri) 
~\psi_{_{\rm ES}}^\star \le(x_i; t'\ri) ~=~ 2 \int \prod_{i=1}^n dx_i \le[x'^T \L x\ri] 
\psi_{_{\rm GS}} \le(x_i; t\ri) ~\psi_{_{\rm GS}}^\star \le(x_i; t'\ri) ,
\eea
where $\L$ is a $N \times N$ matrix given by
\be \la{lambda}
\L =  U^T~K_D^{1/4}~\a~\a^T~K_D^{1/4}~U \equiv
\le( 
\begin{array}{cc} 
\L_A & \L_B \\
\L_B^T & \L_C 
\end{array}
\ri) \, ,
\ee
$\L_A$ is an $n \times n$ symmetric matrix; $\L_B$ is an $n \times
(N-n)$ matrix; $\L_C$ is an $(N-n) \times (N-n)$ symmetric matrix. 
Defining two $(N-n) \times (N-n)$ square matrices $\L_\b$ and $\L_\c$ such that
\bea \la{lambdabetgam}
\L_\b &=& \fr 1 {\k} \le(2 \L_C ~-~ \L_B^T A^{-1} B ~-~ B^T A^{-1} \L_B ~+~ 
B^T A^{-1} \L_A A^{-1} B\ri) \nn\\
\L_\c &=& \fr 1 {\k} \le(2 \L_B^T A^{-1} B ~-~ B^T A^{-1} \L_A 
A^{-1} B\ri) ,
\eea 
one can express the above density matrix for ES in terms of the density
matrix for GS as \cite{sdshankiES}:
\be \label{es-den1}
\r_{_{\rm ES}} (t; t') ~=~ \fr 1 {\k} \le[1 - \fr {t^T \L_\c t + t'^T \L_\c t'} 
2 + t^T \L_\b t'\ri] \r_{_{\rm GS}} (t; t') \, ,
\ee
where $\k = \mbox{Tr} (\L_A A^{-1})$. In the above, the matrix $\L_\b$ is symmetric, 
whereas the matrix $\L_\c$ is not necessarily symmetric due to the presence of the 
first term in the parentheses on the right hand side.

Eq. (\ref{es-den1}) is an exact expression for the density matrix for a discretized 
scalar field with any one HO in the first ES and the rest in the GS. However, unlike 
the GS density matrix $\r_{_{\rm GS}}$, given by Eq. (\ref{gs-den}), the ES density 
matrix $\r_{_{\rm ES}}$, Eq. (\ref{es-den1}), contains non-exponential terms and
hence can not be written as a product of $(N-n)$, $2$-coupled HO density matrices. 
Therefore, the ES entropy cannot be expressed as a sum of $2$-HO entropies, as in the 
case of GS. One may however note that the GS density matrix $\r_{_{\rm GS}}$, Eq. 
(\ref{gs-den}), is a Gaussian that attenuates virtually to zero beyond its few sigma 
limits. Therefore, if
\bea \la{smallness}
\e_1 \equiv t_{max}^T  \L_\b  t_{max}  \ll 1 \quad, \quad 
\e_2 \equiv t_{max}^T  \L_\c  t_{max}~ \ll 1 \,
\eea
where 
\be \la{tmax}
t_{max}^T = \le(\fr{3 (N-n)}{\sq{2 \mbox{Tr}(\c - \b)}} \ri)
\le(1,1,\dots \ri)~
\ee
corresponding to $3\sigma$ limits of the Gaussian inside $\r_{_{\rm GS}}$, then
one may approximate  
\bea
1 &-& \fr {t^T \L_\c t + t'^T \L_\c t'} 2 + t^T \L_\b t' ~\approx~ \exp 
\le[- \fr {t^T \L_\c t + t'^T \L_\c t'} 2 + t^T \L_\b t' \ri].
\eea
Consequently, with a shift of parameters: $\b' \equiv \b + \L_\b , \c' \equiv \c + 
\L_\c$, the ES density matrix above can also be approximated as a Gaussian
\bea \la{es-den2}
\rho_{_{\rm ES}} (t; t') \approx \fr 1 {\k} \exp \le[-\fr {t^T \c' t 
+ t'^T \c' t'} 2 + t^T \b' t'\ri].
\eea
Factorizing this once again into $N-n$ two-HO density matrices, albeit in terms of
the shifted parameters $(\b',\c')$, one can evaluate the ES entanglement entropy 
$S_{_{\rm ES}}$. 

%
\begin{wrapfigure}{r}{0.58\textwidth}
  \vspace{-5pt}
  \begin{center}
    \includegraphics[width=0.55\textwidth,height=91mm]{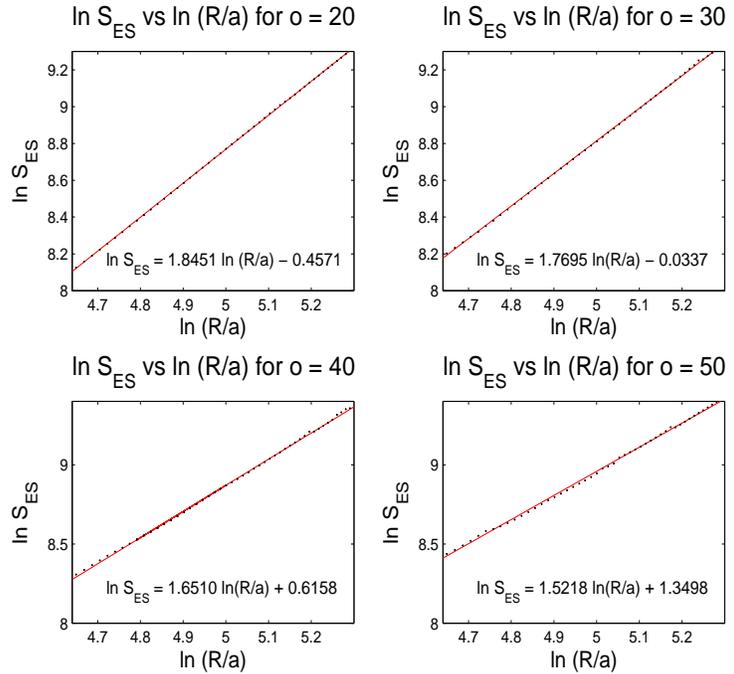}
  \end{center}
  \vspace{-2pt}
  \caption{\small Plot of logarithm of the ES entropy versus $\ln(R/a)$ for $N = 300, ~ n = 
100 - 200$ for different values of $o = 20, 30, 40, 50$. The data approximately fit to straight lines 
with slope $< 2$.}
\la{fig:ES}
\end{wrapfigure}
The entropy computation is done numerically (using MATLAB) in \cite{sdshankiES}, 
with a precision setting $Pr = 0.1 \%$ for the set of parametric values: $N = 300, 
~n = 100 - 200, ~o = 10 - 50$, where $o$ is the number of last non-vanishing entries 
in the vector $\a^T$, i.e., $\a^T = \le(1/\sq{o}\ri) (0, \cdots, 0; 1, \cdots, 1)$. 
The criteria (\ref{smallness}) are found to be satisfied for these choices of the 
parameters. The $\ln S_{_{\rm ES}}$ vs. $\ln (R/a)$ data for different fixed values 
of the amount of excitation $o$ ($= 20, 30, 40, 50$) fit approximately to straight 
lines as shown in Fig. \ref{fig:ES}. The slopes of all these straight lines are 
less than $2$ and the higher the value of $o$ the smaller is the slope. This shows 
that the ES entropy approximately scales as a power of the area:
\be \la{es-ent1}
S_{_{\rm ES}} \simeq \l_0 \le(\fr R a\ri)^{2 \m}\!\!= \fr{\l_0}{4 \pi} 
\le(\frac{\cA}{a^2}\ri)^\m\!; ~ \cA = 4 \pi R^2 .
\ee
The power $\m$, however, is always less than unity and decreases with the increase
in the number of excitations $o$. The coefficient $\l_0$ on the other hand increases
with $o$. Thus contrary to the cases of GS, GCS and SS, the AL is always violated 
in the case of the 1-particle ES \cite{sdshankiES}. 

Now, it is quite interesting to see that even for a small amount of excitation 
($o \sim 20$) the entropy-area relationship is so drastically changed that the AL 
could not be recovered in any limit. Further studies with computations of higher 
precision ($Pr = 0.01 \%$) \cite{sdshankiss} however reveal that there are slight 
variations in the linear fits of $\ln S_{_{\rm ES}}$ vs. $\ln (R/a)$ data for 
different values of $o$. In fact, it has been shown that the ES entropy actually 
approaches the GS entropy (and hence obeys the AL) for very large area 
\cite{sdshankiss}. Thus the excitations seem to give rise to some corrections to 
the AL which are significant for smaller areas, but become negligible for very 
large areas. These corrections are not manifested by the linear fits of the data 
(shown in Fig. \ref{fig:ES}). Therefore a more accurate non-linear fitting is 
required. We will discuss about these corrections to the AL in the next section 
where we consider the scalar field to be in a more general state --- the MS --- 
which is a linear superposition of GS and 1-particle ES.

\subsection{Superposition of ground and excited state} 
\la{ent-ms}

The superposition of ground and excited state wave-function 
$\psi_{_{\rm MS}}$ is given by
\be \la{ms-wavefn}
\psi_{_{\rm MS}} (\xh; t) ~=~ \le[c_0 ~\psi_{_{\rm GS}} (\xh; t) ~+~ 
c_1 ~\psi_{_{\rm ES}} (\xh; t)\ri] \, , 
\ee
where $\psi_{_{\rm GS}}$ is the GS wave-function, given by Eq. (\ref{gs-wavefn}),
$\psi_{_{\rm ES}}$ is the ES wave-function, given by Eq. (\ref{es-wavefn}), 
$\xh \equiv \{x_1, \cdots, x_n\}~$, and as before $t_j \equiv x_{n+j} 
~ (j = 1, \cdots, N-n) ~;~ t \equiv \{t_1, \cdots, t_{N-n}\} = \{x_{n+1}, 
\cdots, x_N\}$. We assume that $c_0$ and $c_1$ are real constants, and 
$\psi_{_{\rm MS}}$ is normalized so that $c_0^2 + c_1^2 = 1$. Using Eq. 
(\ref{es-wavefn}), we can write,
\be \la{ms-wavefn1}
\psi_{_{\rm MS}} (\xh; t) ~=~ \le[c_0 ~+~ c_1 ~ f (\xh; t)\ri] \psi_{_{\rm GS}} 
(\xh; t) \quad , \quad f (\xh; t) ~=~ \sq{2} \a^T K_D^{1/4} U x ~=~ y^T x \, ,
\ee
where the column vector $\a$ includes the expansion coefficients defined in the 
previous section [$\a^T = (\a_1,\dots,\a_N) = (1/\sq{o}) (0, \dots, 0; 1, \dots, 
1)$], and $y$ is an $N$-dimensional column vector $y$ defined as
\bea \la{y}
y ~=~ \sq{2} U^T K_D^{1/4} \a ~=~ \le( \begin{array}{l} 
{y_A} \\ 
{y_B} 
\end{array} \ri) 
\eea
$y_A$ and $y_B$ are $n$- and $(N-n)$-dimensional column vectors, respectively.

The MS density matrix can be expressed as a sum of three terms:
\bea \la{ms-den}
\r_{_{\rm MS}} (t; t') ~=~ \int \prod_{i=1}^n dx_i~ \psi_{_{\rm MS}} (\xh; t) 
\psi_{_{\rm MS}}^\star (\xh; t') ~=~ c_0^2 ~\r_{_{\rm GS}} (t; t') ~+~ c_1^2 
~\r_{_{\rm ES}} (t; t') ~+~ c_0 c_1 ~\r_{_{\rm X}} (t; t') \, ,
\eea
where $\r_{_{\rm GS}} (t; t')$ is the GS density matrix, Eq. (\ref{gs-den}), 
$\r_{_{\rm ES}} (t; t')$ is the ES density matrix, Eq. (\ref{es-den}), and
$\r_{_{\rm X}} (t; t')$ is a cross term due to the superposition of GS and
ES. Identifying the matrix $\Lambda$, its components, and the constant $\k$ 
(defined for ES in the previous section), with the column vector $y$ and its 
components:
\bea \la{identify}
&& \L ~=~ \fr 1 2 y y^T ~=~ \le( \begin{array}{ll} 
{\L_A} & {\L_B} \\
{\L_B^T} & {\L_C} 
\end{array} \ri), \nn\\
&& \L_A = \fr 1 2 y_A y_A^T ~;~ \L_B = \fr 1 2 y_A y_B^T ~;~ 
\L_C = \fr 1 2 y_B y_B^T ~, \nn\\
&& \k ~=~ \mbox{Tr} (\L_A A^{-1}) ~=~ \fr 1 2 y_A^T A^{-1} y_A \, ,
\eea
one can evaluate the cross term $\r_{_{\rm X}}$ as 
\bea \la{cross-den}
\r_{_{\rm X}} (t; t') ~=~ \int \prod_{i=1}^n dx_i \le[f (\xh; t) + f (\xh; t')\ri] 
\psi_{_{\rm GS}} (\xh; t) ~ \psi_{_{\rm GS}}^\star (\xh; t') ~=~ \le(y_B - p\ri)^T 
\le(t + t'\ri) \r_{_{\rm GS}} (t; t') \, ,
\eea
where $p$ is an $(N-n)$-dimensional column vector defined by
\be \la{p}
p ~=~ B^T A^{-1} y_A \, .
\ee

The full density matrix for the MS, Eq. (\ref{ms-den}), thus reduces to
\be \la{ms-den1}
\r_{_{\rm MS}} (t; t') = \le[c_0^2 + c_1^2 \k \le\{1 + u (t; t')\ri\} + c_0 c_1  
v (t; t')\ri] \r_{_{\rm GS}} (t; t') \, ,
\ee
where 
\bea \la{uv}
u (t; t') ~=~ - ~ \fr{t^T \L_\c t + t'^T \L_\c t'} 2 ~+~ t^T \L_\b t' \quad, 
\quad v (t; t') ~=~ \le(y_B - p\ri)^T \le(t + t'\ri) \, . 
\eea
Defining
\be \la{F}
F (t; t') ~=~ 1 ~+~ \k_1 w (t; t') ~+~ \k_2 v (t; t') ~+~ \fr {\k_2^2} 2 v^2 
(t; t') \, ,
\ee
where
\bea 
\la{w}
&& w (t; t') ~=~ - ~ \fr{t^T \L_{\c'} t + t'^T \L_{\c'} t'} 2 ~+~ 
t^T \L_{\b'} t' \, ,  \nn \\ 
\L_{\b'} &=& \L_\b ~-~ 2 \k_0 \le(\L_\b ~-~ \fr {\L_C}{\k}\ri) \quad, \quad
\L_{\c'} ~=~ \L_\c ~+~ 2 \k_0 \le(\L_\b ~-~ \fr {\L_C}{\k}\ri) \, , \nn \\
\k_0 &=& \fr{c_0^2}{\kt} \quad ; \quad \k_1 ~=~ \fr{c_1^2}{\kt} \quad ; \quad
\k_2 ~=~ \fr{c_0 c_1}{\kt} \quad , \quad \quad \kt ~=~ c_0^2 + c_1^2 \k \, ,
\eea
the density matrix (\ref{ms-den1}) can be written as
\be \la{ms-den2}
\r_{_{\rm MS}} (t; t') ~=~ \kt ~ F (t; t') ~ \r_{_{\rm GS}} (t; t') \, .
\ee
In the above, $\L_{\b'}$ and $\L_{\c'}$ are $(N-n) \times (N-n)$ matrices, and
constants $(\k_0, \k_1, \k_2)$ describe the amount of mixing between the GS and ES.

Now, similar to the case of ES, here also the pre-factor $F (t; t')$ of the Gaussian 
$\r_{_{\rm GS}} (t; t')$ in Eq. (\ref{ms-den2}) contains non-exponential terms. 
Therefore $\r_{_{\rm MS}}$ cannot be factorized into $(N-n)$ two-HO density matrices. 
However, as mentioned in sec. \ref{ent-es1}, that when the vector $t^T$ is outside the 
maximum $t_{max}^T$, Eq. (\ref{tmax}), corresponding to the $3 \sigma$ limits, the 
Gaussian inside $\r_{_{\rm GS}} (t; t')$ is negligible. Therefore, if the conditions 
(\ref{smallness}) (given in sec. \ref{ent-es1}) as well as the conditions
\bea \la{smallness1}
\et_1 \equiv t_{max}^T  \L_{\b'}  t_{max}  \ll 1~,~ 
\et_2 \equiv t_{max}^T  \L_{\c'}  t_{max}~ \ll 1 
\eea
are satisfied, then keeping terms up to quadratic order in $t, t'$, the pre-factor 
$F (t; t')$ can be approximated as 
\be \la{F-approx}
F (t; t') ~\approx~ \exp \le[\k_1 ~w (t; t') ~+~ \k_2 ~v (t; t')\ri] \, .
\ee
Then by using Eq.(\ref{gs-den}) for $\r_{_{\rm GS}} (t; t')$ the (approximated) MS 
density matrix can be expressed in the form:
\be \la{ms-den3}
\r_{_{\rm MS}} (t; t') ~=~ \kt \sq{\fr{|\O|}{\pi^{N-n} |A|}} ~ \exp \le[z (t; t') + 
\k_2 v (t; t')\ri] \quad , \quad z (t; t') = -~ \fr{t^T \c' t + t'^T \c' t'} 2 + 
t^T \b' t' \, , 
\ee
where the $(N-n) \times (N-n)$ matrices $\b'$ and $\c'$ are defined by
\bea
\la{betagammapr}
\hspace*{-0.6cm} 
\b' &=& \b + \k_1 \L_{\b'} =\b + \k_1 \L_\b - 2 \k_0 \k_1 
\le(\L_\b - \frac{\L_C}{\k}\ri) \nn\\
\hspace*{-0.6cm}
\c' &=& \c + \k_1 \L_{\c'} = \c + \k_1 \L_\c + 2 \k_0 \k_1 
\le(\L_\b - \frac{\L_C}{\k}\ri) \, . 
\eea
The matrix $\b'$ is symmetric while the matrix $\c'$ is not necessarily symmetric.

Now shifting $(N-n)$ variables $t \equiv \{x_{n+1}, \cdots, x_N\}$ and $t' \equiv
\{x'_{n+1}, \cdots, x'_N\}$ by constant values $s \equiv \{s_1,\cdots, s_{N-n}\}$:
\be \la{var-shift}
t ~\rightarrow~ t ~+~ s ~;~~~ t' ~\rightarrow~ t' ~+~ s \, ,
\ee
the density matrix (\ref{ms-den3}) becomes
\be \la{ms-den4}
\r_{_{\rm MS}} (t; t') ~=~ \cN ~\exp \le[- ~ \fr{t^T \c' t + t'^T \c' t'} 2 ~+~ 
t^T \b' t' \ri] \quad, \quad 
\cN = \kt \sq{\fr{|\O|}{\pi^{N-n} |A|}} ~\exp \le[- s^T \le(\b' - \c'\ri)^T s\ri] \, .
\ee
The $(N-n)$-dimensional constant column vector $s$ is determined from the equation
\be \la{s-eq}
s^T \le(\b' - \frac{\c' + \c'^T} 2\ri) = - \k_2 \le(y_B - 
B^T A^{-1} y_A\ri) \, .
\ee
For either $c_0 = 0$ or $c_1 = 0$, the constant $\k_2 = 0$, whence from the above 
equation (\ref{s-eq}), we have $s = 0$. Thus it can be verified that the MS density 
matrix (\ref{ms-den4}) reduces to the GS density matrix (\ref{gs-den}) when $c_0 = 1, 
c_1 = 0$, which imply $\b' = \b, \c' = \c$.  On the other hand when $c_0 = 0, c_1 = 1$, 
which imply $\b' = \b + \L_\b, \c' = \c + \L_\c$, the MS density matrix (\ref{ms-den4}) 
is the same as ES density matrix (\ref{es-den}). In general, when both $c_0$ and $c_1$ 
are non-vanishing, then under the shifts $\b \rightarrow \b', \c \rightarrow \c'$ 
(where $\b'$ and $\c'$ are given by Eqs. (\ref{betagammapr})) the MS density matrix 
(\ref{ms-den4}) is of the same form as the GS density matrix (\ref{gs-den}), up to a
normalization factor given above.  Such a normalization constant does not affect the 
entropy computation.  Therefore the total MS entropy $S_{_{\rm MS}}$ can be evaluated 
following the same steps [Eqs. (\ref{diag}) -- (\ref{gs-ent})] as in the case of GS, 
albeit with the replacements $\b \rightarrow \b', \c \rightarrow \c'$.

Computation of the entanglement entropy has been done numerically (using MATLAB) in
\cite{sdshankiss}, for $N = 300, ~n = 100 - 200, ~o = 30, 40, 50$, and with a precision 
setting $Pr = 0.01 \%$ in each of the following cases: 
\begin{description}
\item (i) GS ($c_0 = 1, c_1 = 0$),
\item (ii) ES ($c_0 = 0, c_1 = 1$), 
\item (iii) an equal mixing (MS$_{Eq}$) of ES with GS ($c_0 = 
c_1 = 1/\sqrt{2}$), and 
\item (iv) a high mixing (MS$_{Hi}$) of ES with GS ($c_0 = 1/2, 
c_1 = \sqrt{3}/2$). 
\end{description}
The conditions (\ref{smallness}) as well as (\ref{smallness1}) are found to be satisfied 
for the above values of the parameters.

Before proceeding to the results we would like to mention the following: the expectation 
value of energy, $\cE$, for MS turns out to be
\be \la{MS-energy}
\cE = \langle \psi_{_{\rm MS}}| H | \psi_{_{\rm MS}} \rangle = \cE_0 + \fr {c_1^2} o 
\sum_{i=N-o+1}^N k_{Di}^{1/2} \, ,
\ee
where $\cE_0 = \fr 1 2 \sum_{i=1}^N k_{Di}^{1/2}$ is the (zero-point) GS energy. The 
fractional excess of energy over the zero-point energy is therefore given by
\be \la{MS-excess}
\fr {\D \cE}{\cE_0} = \fr{\cE - \cE_0}{\cE_0} = \fr {2 c_1^2} o \le[1 + \fr{
\sum_{i=1}^{N-o} k_{Di}^{1/2}}{\sum_{i=N-o+1}^N k_{Di}^{1/2}}\ri]^{-1} \, .
\ee
Now, the value of $c_1$ is between $0$ and $1$ and as mentioned
earlier $k_{Di} > k_{Dj}$ for $i > j$. Therefore even in the extreme
situation $c_1 = 1$, i.e., ES, with a fairly high amount of excitation
$o \sim 50$, the fractional change in energy is at most about $\sim 4
\%$. Moreover, since there are $o$ number of terms in the sum in the
second term of Eq.(\ref{MS-energy}), the excitation energy $(\cE -
\cE_0) \sim 1$ (in units of $1/a$, where $a$ is the lattice
spacing). Hence, if we choose $a \sim$ Planck length, then this
excitation energy is of the order of Planck energy. As the mass of a
semi-classical black hole is much larger than the Planck mass, one may
therefore safely neglect the back-reaction of the scalar field on the
background.

\begin{figure*}[!htb]
\begin{center}
\epsfxsize 5.5 in
\epsfysize 2.25 in
\epsfbox{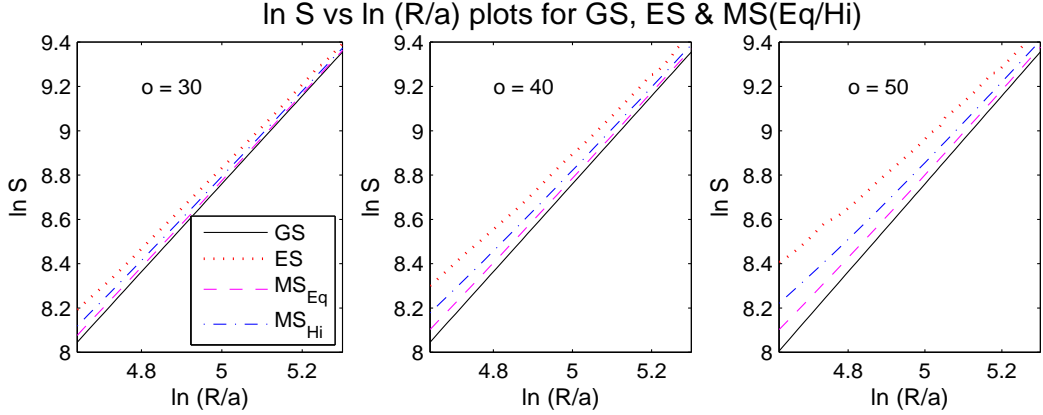}
\caption{\small Plots of logarithm of GS, ES and MS (Eq/Hi) entropies versus $\ln(R/a)$ for 
$N = 300, ~ n = 100 - 200$ and $o = 30, 40, 50$. The numerical error in the computation
is less than $0.01\%$.}
\la{fig:MS}
\end{center}
\vspace*{-0.05cm}
\end{figure*}
%
\begin{figure*}[!htb]
\begin{center}
\epsfxsize 5.5 in
\epsfysize 2.25 in
\epsfbox{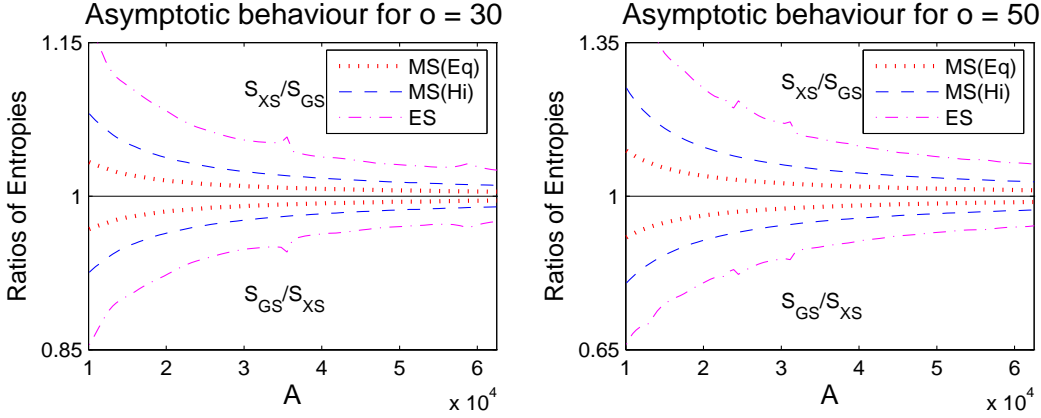}
\caption{\small Ratios of GS and MS (Eq/Hi) or ES entropies and their reciprocals plotted
against the area $\cA$ (in units of $a^2$) for $o = 30, 50$, to show the asymptotic 
nature of the MS and ES entropies with respect to the GS entropy. The curves on the 
upper half (above $1$) show the variation of $S_{_{\rm XS}}/S_{_{\rm GS}}$ with $\cA$, 
where XS stands for MS(Eq/Hi) or ES, while the lower curves show the variation of 
$S_{_{\rm GS}}/S_{_{\rm XS}}$ with $\cA$.}
\la{fig:MS-asymp}
\end{center}
\vspace*{-0.55cm}
\end{figure*}

Fig. \ref{fig:MS} shows the plots of the logarithm of the total entropy
$S$ vs. $\ln (R/a) = \ln (n + 1/2)$, for the cases of GS, ES and MS (Eq/Hi)
with different values of the excitation ($o = 30, 40, 50$).  The plot for
GS is the linear fit with slope $\simeq 2$ shown earlier.  The plots for
the MS (Eq/Hi) cases, as well as for ES, are nearly linear for different 
values of the excitations $o = 30, 40, 50$ and appear to coincide with the 
plot for GS for large areas ($\cA = 4 \pi R^2 \gg a^2$). For a closer 
examination of this, the ratios $S_{_{\rm MS}} ({\rm Eq or Hi})/S_{_{\rm GS}}, 
S_{_{\rm ES}}/S_{_{\rm GS}}$ and their inverse are plotted against the area 
$\cA$ in Fig. \ref{fig:MS-asymp}. All these ratios approach to unity with 
increasing area for different excitations ($o = 30, 50$), i.e., the MS (Eq/Hi) 
and the ES entropies coincide asymptotically with the GS entropy, following the
criterion of `asymptotic equivalence' \cite{asymp}:
\be \la{asymp-eq}
\mbox{lim}_{_{\cA \rightarrow \infty}} \fr {S_{_{\rm XS}} (\cA)} {S_{_{\rm GS}} 
(\cA)} = 1 \quad ; \quad \mbox{lim}_{_{\cA \rightarrow \infty}} \fr {S_{_{\rm GS}} 
(\cA)} {S_{_{\rm XS}} (\cA)} = 1 \quad , \quad \quad {\rm XS} \equiv {\rm MS~ (Eq~ 
or~ Hi)~ or~ ES} \, .
\ee
From Fig. \ref{fig:MS-asymp} one can also observe that the MS(Eq) entropy is closer 
to the GS entropy for large $\cA$, than the MS(Hi) entropy and the ES entropy, the 
latter being the farthest. Thus for smaller values of the relative weight $c_1$ of 
the mixing of ES with GS the asymptote is sharper.

%
\begin{figure*}[!htb]
\begin{center}
\epsfxsize 5.5 in
\epsfysize 2.25 in
\epsfbox{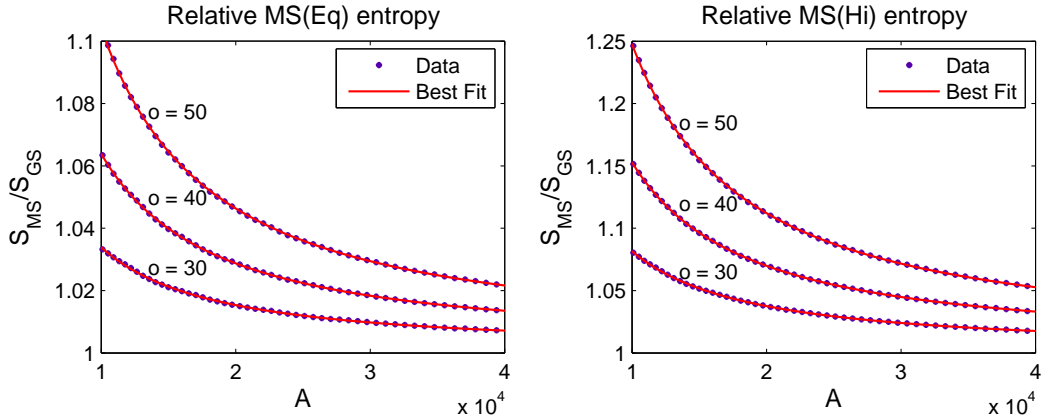}
\caption{\small Best fit plots (solid lines) of the relative mixed state entropies 
($S_{_{\rm MS}}/S_{_{\rm GS}}$) for equal and high mixings versus the area $\cA$ 
(in units of $a^2$), for $o = 30, 40, 50$. The corresponding data are shown by 
asterisks.}
\la{fig:MS-fit}
\end{center}
\vspace*{-0.55cm}
\end{figure*}
%

\begin{table*}[!htb]
\la{table1}
\begin{center}
\begin{tabular}{|c||c|c|c||c|c|c|}
\hline
Fitting&\multicolumn{3}{c|}{For MS$_{Eq}$}&\multicolumn{3}{c|}{For MS$_{Hi}$} \\
\cline{2-7}
Parameters&$~o = 30~$&$~o = 40~$&$~o = 50~$&$~o = 30~$&$~o = 40~$&$~o = 50~$ \\
\hline\hline 
& & & & & &  \\
${\tilde \s}_0$&$~1.001~$&$~1.002~$&$~1.003~$&$~1.001~$&$~1.004~$&$~1.006~$ \\
& & & & & &  \\
${\tilde \s}_1$&$~1738~$&$~4288~$&$~8039~$&$~2956~$&$~7652~$&$~14120~$ \\  
& & & & & &  \\
${\tilde \n}$&$~1.180~$&$~1.210~$&$~1.225~$&$~1.141~$&$~1.178~$&$~1.192~$ \\
\hline
\end{tabular}
\end{center}
\caption{\small Values of the parameters of the fit $S_{_{\rm MS}}/S_{_{\rm GS}} = 
{\tilde \s}_0 + {\tilde \s}_1 \le(\cA/a^2\ri)^{- {\tilde \n}} $ for both MS(Eq) and 
MS(Hi) cases with excitation $o = 30, 40, 50$.}
\end{table*}

Fig. \ref{fig:MS-fit} shows the best fit ratios of the MS entropies (for equal and high 
mixings, with $o = 30, 40, 50$) to the GS entropy, which follow a simple formula:
\be \la{MS-fita}
\frac{S_{_{\rm MS}}}{S_{_{GS}}} =  {\tilde \s}_0 ~+~ {\tilde \s}_1 
\le(\fr{\cA}{a^2}\ri)^{- {\tilde \n}} \, ,
\ee
where the values of the fitting parameters ${\tilde \s}_0, {\tilde \s}_1$ and ${\tilde \n}$ 
are shown in Table 1 for different values of $o = 30, 40, 50$. For all these values of $o$,
the parameter ${\tilde \s}_0 \approx 1$ in both MS(Eq) and MS(Hi) cases. The parameter 
${\tilde \s}_1$ is of the order of $10^3$ and increases with increasing excitations. The
parameter ${\tilde \n}$ lies between $1$ and $1.25$ for the above values of $o$, and also
increases with increasing $o$. Using the expression for the GS entropy, viz., $S_{_{\rm GS}} 
= n_0 (\cA/a^2)$, where $n_0$ is a constant, we can rewrite the above Eq. (\ref{MS-fita}) as
\be \la{MS-fit}
S_{_{\rm MS}} ~=~ \s_0 \le(\fr{\cA}{a^2}\ri) + \s_1 \le(\fr{\cA}{a^2}\ri)^{-\n} \, ,
\ee
where $\s_0 = n_0 {\tilde \s}_0, \s_1 = n_0 {\tilde \s}_1 \propto c_1$ and $\n = {\tilde \n} - 1$. The
exponent $- \n$ lies between $0$ and $- 0.25$ for both equal and high mixings with the
above values of $o$. It is instructive to stress the implications of the above
result: \\
(i) For the pure vacuum wave-functional, $c_ 1 = 0$ and $\Sen$ is identical to Bekenstein-Hawking
entropy. This clearly shows that the entanglement entropy of 
ground state leads to the area law and the excited states contribute 
to the power-law corrections. \\
(ii) For large black-holes, power-law correction falls off rapidly
and we recover $\SBH$. However, for the small black-holes, the second
term dominates and black-hole entropy is no more proportional to area.
Physical interpretation of this result is immediately apparent. In the
large black-hole (or low-energy) limit, it is difficult to excite the
modes and hence, the ground state modes contribute significantly to
$\Sen$. However, in the small black-hole (or high-energy) limit,
larger number of field modes can be excited and hence they contribute
significantly to $\Sen$. \\
(iii) The power-law corrections to the Bekenstein-Hawking area law derived here in the context
of entanglement of scalar fields have features similar to those derived in the case
of brick-wall model \cite{Sarkar:2007uz} and higher-derivative gravity \cite{Wald:1993a}. 
For instance, it was shown that the entropy of five-dimensional Boulware-Deser 
black-hole \cite{HDentropy} is given by
\bea
S ~=~ \fr{\cA}{4} ~+~ c~ \cA^{1/3} \qquad ;~~~~
c = \mbox{~constant} \, .
\eea
As in Eq. (\ref{MS-fit}) the above entropy is proportional to area for
large horizon radius, however it strongly deviates in the small
horizon limit. It is important to note that the corrections to the
black-hole entropy are generic and valid even for black-holes in
General relativity without any higher curvature terms\footnote{In this
context, it should be mentioned that it is not possible to check for
logarithmic corrections to the entropy in our analysis, as the
numerical error we obtain is much larger than $\ln (n + 1/2)$.}.

\section{Location of the degrees of freedom}
\la{sf-dof}

In the previous section, we showed that the entanglement entropy provides generic power-law 
corrections to the area. We also showed that the quantum DOF that contribute to the area 
law and the subleading corrections are different. As we had mentioned in the introduction, 
this leads to another question: {\it To what extend does the quantum DOF close to or 
far from the horizon contribute to the black-hole entropy?} In this section, we address
this question and show that large contribution $(\sim 97 \%)$ to the area-law comes from 
close to the horizon while the subleading contributions has a larger contribution 
from the regions far from the horizon. 

Let us recall the expression for the interaction matrix $K$, with elements $K_{ij}$ given by 
Eq. (\ref{kij}), for the system of $N$ HOs. The last two terms which signify the 
nearest-neighbour (NN) interaction between the oscillators, are solely responsible for 
the entanglement entropy, i.e., if these two terms are set to zero the entropy vanishes. 
In order to find which DOF give rise to the entropy or what are their contributions, let 
us perform the following operations on the matrix $K$ \cite{sdshankiDoF}:

\bigskip
\noindent
\underline{\bf Operation 1:}
\bigskip

\begin{figure*}[!htb]
\begin{center}
\includegraphics[width=3in,height=2.7in]{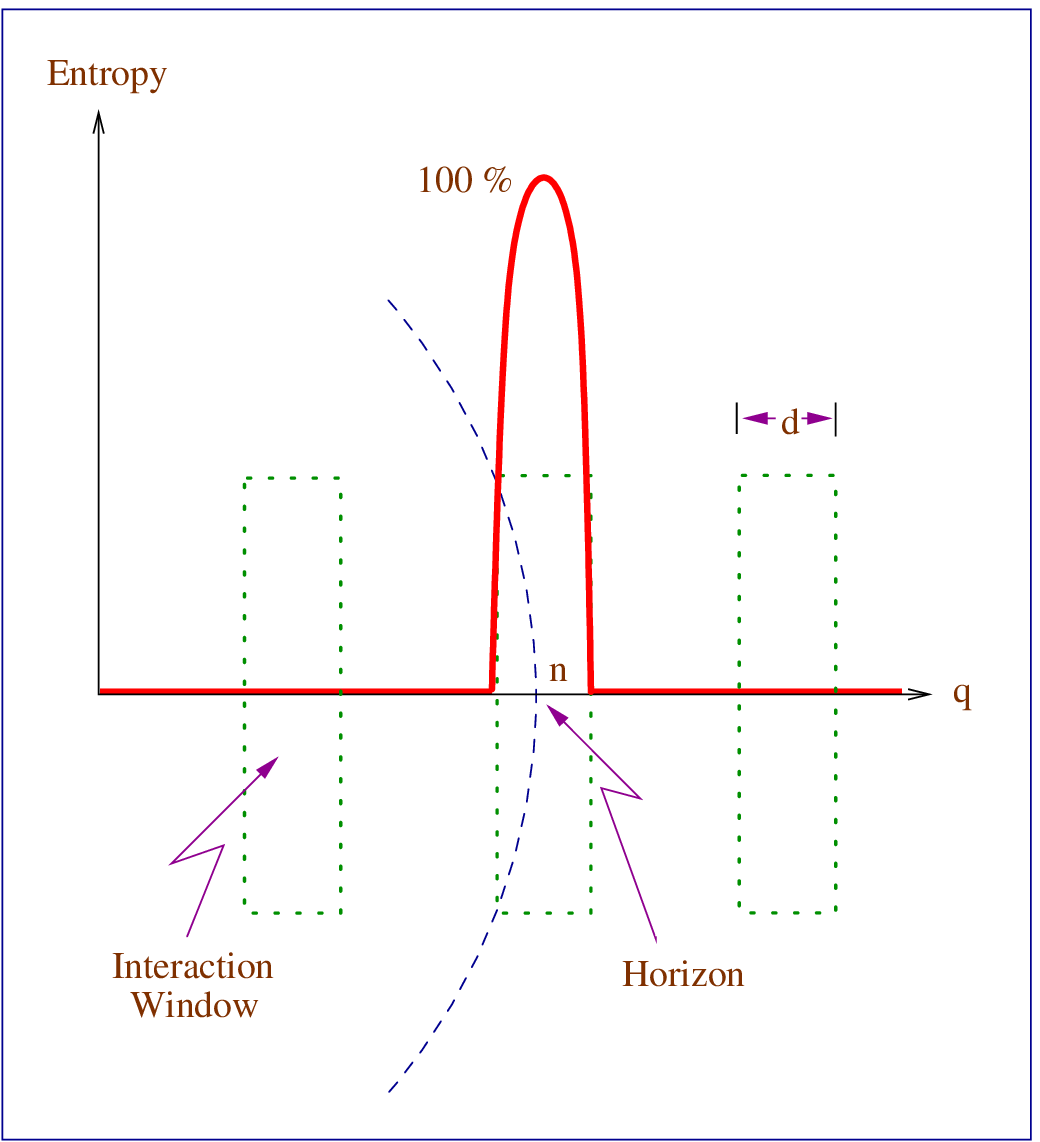}
\includegraphics[width=3.25in,height=2.75in]{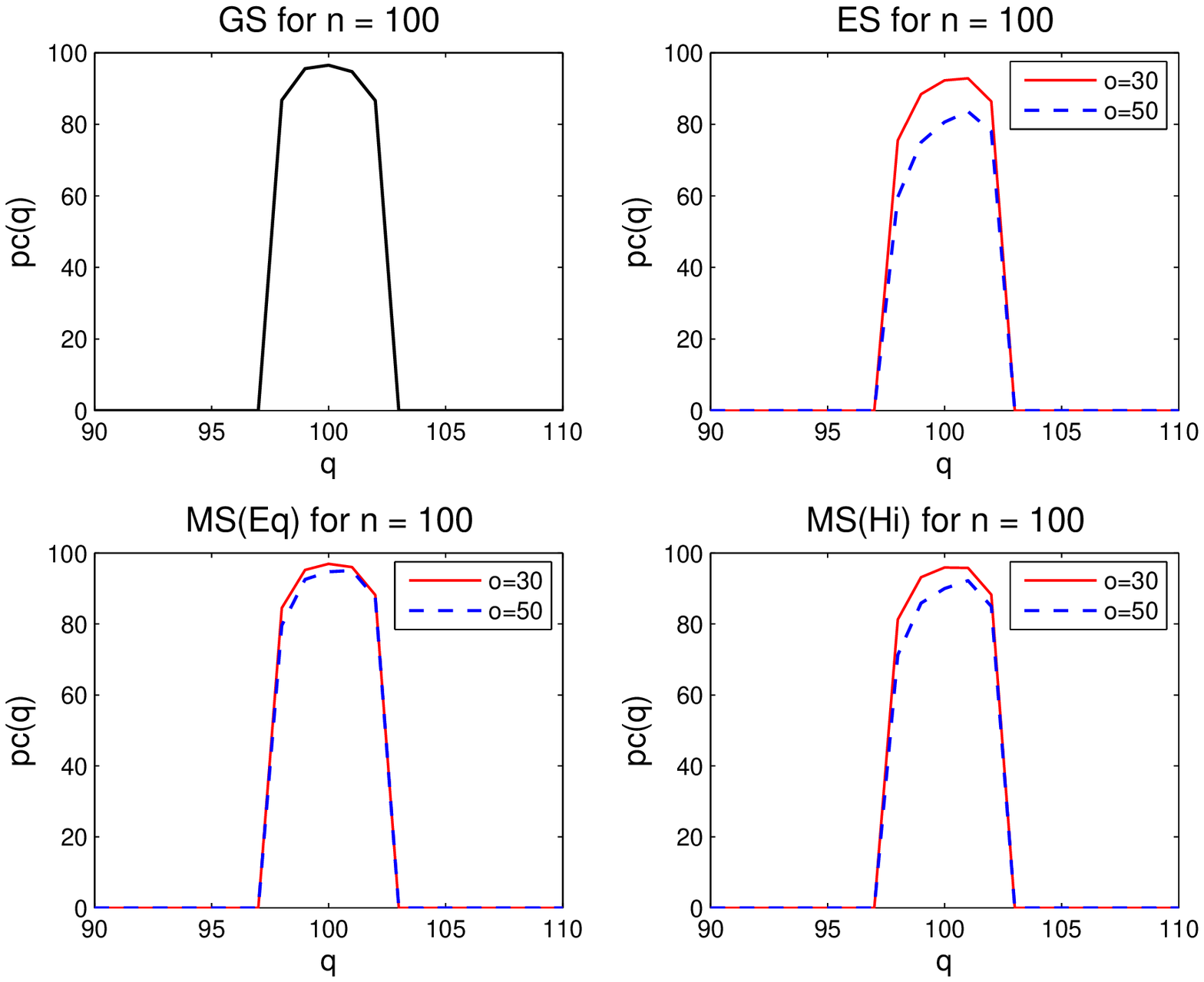}
\caption{\small The figure on the left is a pictorial representation of the entropy due 
to changing positions of the interaction window (green dotted boxes). Contributions to
the entropy are observed only when the window includes the horizon at $n$. The figures
on the right are the actual plots of the percentage contribution $pc(q)$ to the total 
entropy as a function of window position $q$, for a fixed window size $d = 5$ and
fixed $N = 300, n = 100$, in each of cases of GS, ES and MS (Eq/Hi). For ES and MS 
(Eq/Hi) the solid curve is for $o = 30$ whereas the broken curve is for $o = 50$.}
\label{fig:DOF-vary-q}
\end{center}
\vspace*{-0.55cm}
\end{figure*}

Let us set the off-diagonal elements of $K$, which signify the NN interactions, 
to zero (by hand) everywhere except in a `window', whose center is at a point $q$.
The indices $i,j$ of the matrix $K$ run from $q - s$ to $q + s$, where $s \leq q$,
so that the interaction region is restricted to a width of $d = 2s + 1$ radial
lattice points. For instance, with $s = 1$ the window is of size $3 \times 3$, and
the matrix $K$ is schematically depicted as:
\begin{center}
\be
K  = \le( \begin{array}{lllllll} 
{\times}  & {}  & {}& {} & {} & {} & {}  \\
{} & {\times} & {} & {} & {} & {}  & {} \\
{}  & {} & {\times} & {} & {} & {} & {}   \\
{} & {}  & {} & | \overline{{\times}} & \overline{{\times}} & \overline{{}}~~| & {}  \\
{} & {} & {}  & | {{\times}} & {{\times}} & {{\times}}| & {}  \\
{} & {} & {} & |\underline{\null} {}  & \underline{{\times}} & \underline{{\times}} |  & {}  \\
{} & {} & {} & {} & {}  & {} & {\times}  \\
\end{array} \ri) \, .
\ee
\end{center}
Now we let the center $q$ of window to vary from $0$ to a value greater than $n$, 
and thus allow the window to move rigidly across from the origin to a point outside 
the horizon at $n$ as shown pictorially in Fig. \ref{fig:DOF-vary-q}. For every window 
location (values of $q$) and fixed window width $d$, we compute the entanglement
entropy $S (q, {\rm fixed~} d)$ in each of the cases GS, ES and MS (Eq/Hi) and find 
the percentage contribution of the entropy as a function of $q$:
\be \la{pc-vary-q}
pc (q) ~=~ \fr{S (q, {\rm fixed~} d)}{S_{\rm total}} \times 100 \, ,
\ee
where $S_{\rm total}$ is the total entropy with all NN interactions, i.e., with the 
indices $i,j$ running from $0$ to $N$. The variations of $pc (q)$ with $q$ for a window 
width of $d=5$ lattice points is shown in Fig. \ref{fig:DOF-vary-q}, for fixed values 
$N = 300, n = 100$ in each of the cases GS and ES, MS (Eq/Hi) with $o = 30, 50$. 

%

In all the cases of GS, ES and MS (Eq/Hi) there is no contribution to the entropy,
i.e., $pc (q) = 0$, when the interaction window does not include the horizon at $n$. 
The contributions to the entropy rise significantly when the window includes the
horizon, i.e., when the window center $q$ is very close to $n$. In the case of GS, 
$pc (q)$ peaks when the window is symmetrically placed between inside and outside,
i.e., when $q = n$, and decreases when $q$ moves away from $n$. In the ES and MS(Eq/Hi)
cases, however, the peak tends to shift towards a value $q > n$ and the amplitude of
the peak diminish with increasing $o$ and/or the mixing weight $c_1$. As a result, the 
entire profile of $pc (q)$ is more and more asymmetric with increasing $o$ and/or $c_1$,
when the window is symmetrically placed between inside and outside of the horizon. The 
peak is shortest and the profile is most asymmetric for ES with $o = 50$, as shown in 
Fig. \ref{fig:DOF-vary-q}. On the whole the above results thus confirm that the 
entanglement between the scalar field DOF inside and outside the horizon gives rise 
to the entropy, and the DOF in the vicinity (inside or outside) of the horizon 
contribute most to the total entropy while the DOF that are far from the horizon
contribute a small portion that remains. Such contributions from the far-away DOF 
increase with increasing excitations and/or amount of mixing of ES with GS. This is 
indicated by the diminishing maxima of $pc (q)$ in the MS(Eq/Hi) and ES cases, 
compared to the case of GS. To estimate the contribution of the far-away DOF on the 
total entropy we perform an alternate operation on the interaction matrix $K$ as 
described below.

\begin{figure*}[!htb]
\begin{center}
\includegraphics[width=3in,height=2.7in]{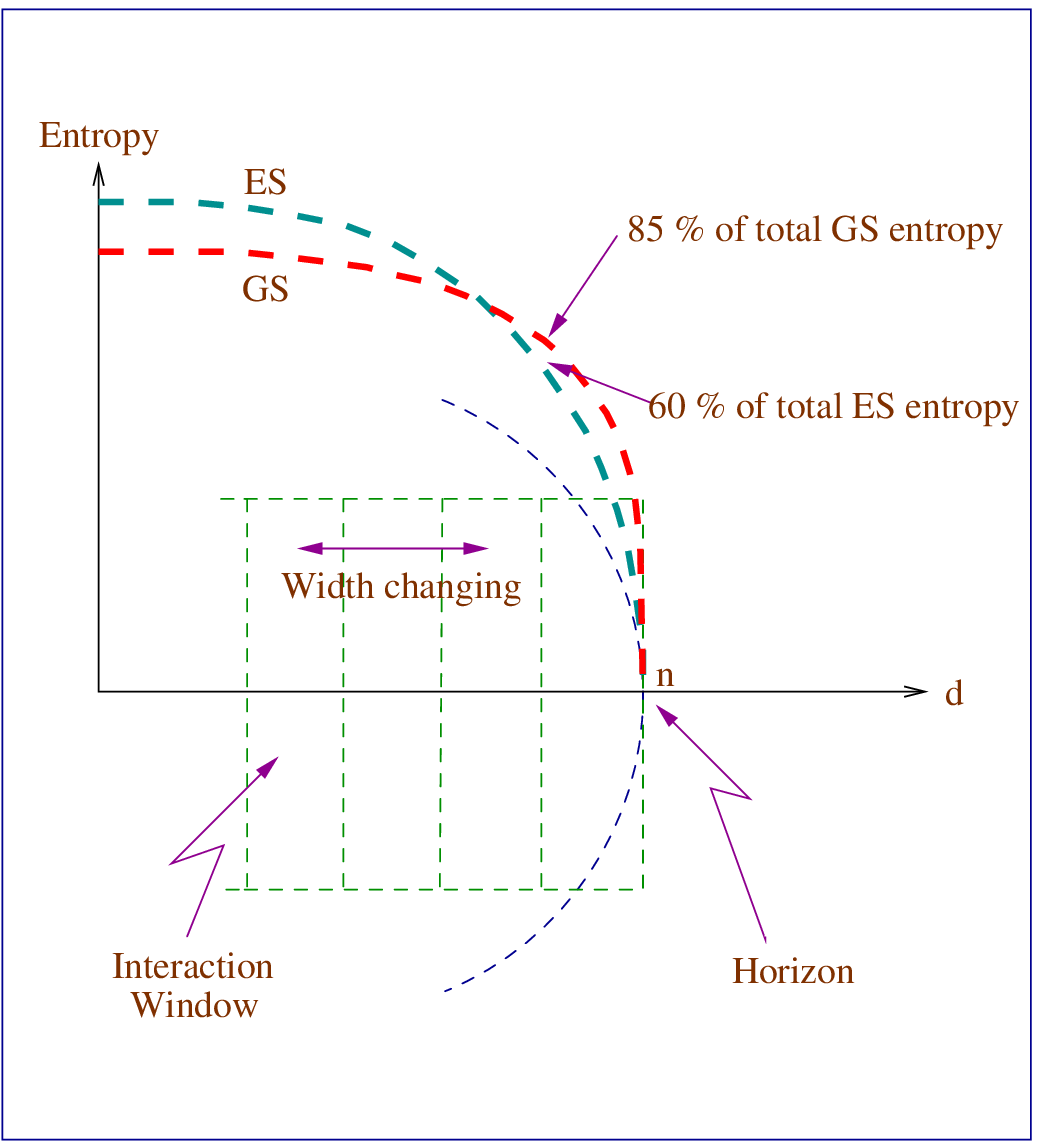}
\includegraphics[width=3.25in,height=2.75in]{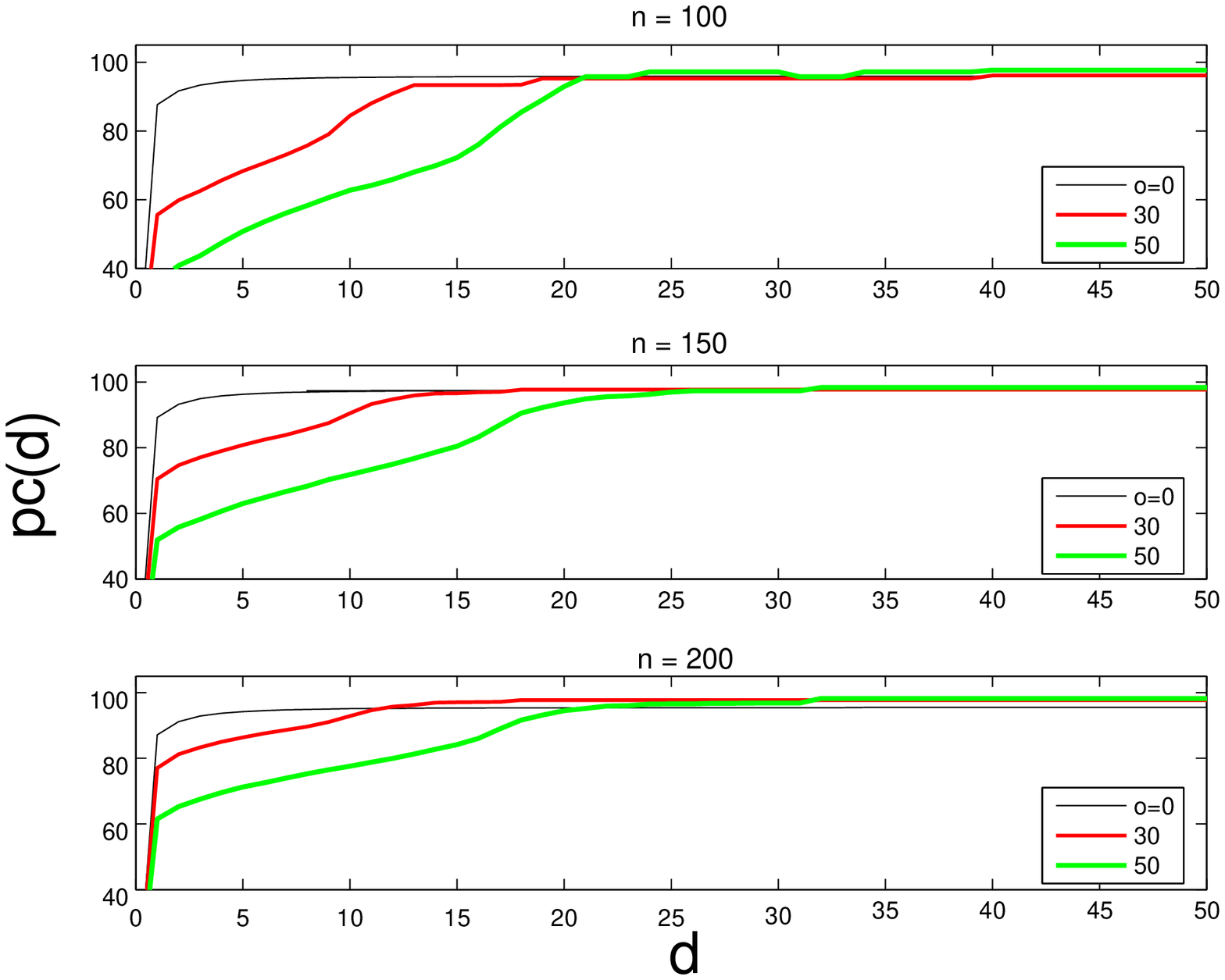}
\caption{\small The figure on the left pictorially shows how the contribution to the
total entropy increases with increasing width $d$ of the interaction window (green 
dashed box) in the cases of GS and ES. The figures on the right are the actual plots 
of the percentage contribution $pc(d)$ to the total entropy as a function of $d$, for 
fixed $N = 300, n = 100, 150, 200$, in each of cases of GS ($o = 0$) and ES (with
$o = 30, 50$). The solid thin curve is for GS ($o = 0$), whereas the bold light and 
thick curves are respectively for ES with $o = 30$ and ES with $o = 50$.}
\label{fig:DOF-vary-d}
\end{center}
\vspace*{-0.55cm}
\end{figure*}

\bigskip
\noindent
\underline{\bf Operation 2:}
\bigskip

let us again set the off-diagonal elements of the matrix $K$ to zero (by hand) 
everywhere except in a `window' whose outer boundary is the horizon at $n$. The
index $i$ of the elements of the matrix $K$ therefore runs from a point $p$ to 
$n$, where $0 \leq p \leq n$. Now we let the width $d = n - p$ of the window to 
vary from $0$ to $n$, as shown pictorially in Fig. \ref{fig:DOF-vary-d}. Computing 
the entanglement entropy $S (d)$ for every window width $d$, one then finds the 
percentage contribution of the entropy as a function of $d$:
\be \la{pc-vary-d}
pc (d) ~=~ \fr{S (d)}{S_{\rm total}} \times 100 \, ,
\ee
where $S_{\rm total}$ is the total entropy which is recovered for the full width
$d = n$, i.e., $p = 0$. For convenience we consider here the two extreme cases: GS 
($o = 0$) and ES (with $o = 30, 50$). The effects of the DOF on the entropy in the 
MS (Eq/Hi) cases are intermediate between those for the GS and ES cases, and are
therefore of not much interest.

%

The variation of the percentage contribution $pc(d)$ with the window width $d$ is shown 
in Fig. \ref{fig:DOF-vary-d} for parametric choices of $n = 100, 150, 200$. In the case 
of GS, almost the entire entropy is recovered within a width of just $d = 3$ . This 
again shows that most of the total GS entropy, which obeys the AL, is contributed 
by the interaction region that encompassed the DOF very close to the horizon. In the 
case of ES, for which the AL is violated, it takes $d$ to be as much as $15 - 20$, 
depending on the excitations $o = 30 - 50$, so that the total ES entropy is recovered.
Therefore the DOF that are far-away from the horizon have a greater contribution to
the entropy in the case of ES, than for GS, and such contributions increase with
increasing excitations $o$. The location of the horizon (values of $n$) are also
found to affect the entropy contributions. In order to estimate of how much the DOF 
that are far from the horizon affect the total entropy in the case of ES, as compared 
to the case of GS, let us consider the percentage increase in entropy for an increment 
in the interaction region by exactly one lattice point, i.e.,   
\be
\D pc (d) = pc(d) - pc(d-1) \, .
\ee
This is given by the slope of the $pc (d)$ vs. $d$ plots shown in Fig. \ref{fig:DOF-vary-d}.
In Fig. \ref{fig:DOF-delta} the variations of $\D pc (d)$ with $(n-d)$ are shown for $n = 
100, 150, 200$ in the cases of GS and ES (with $o = 30, 50$). 

%
\begin{figure*}[!htb]
\begin{center}
\epsfxsize 4.75 in
\epsfysize 3.5 in
\epsfbox{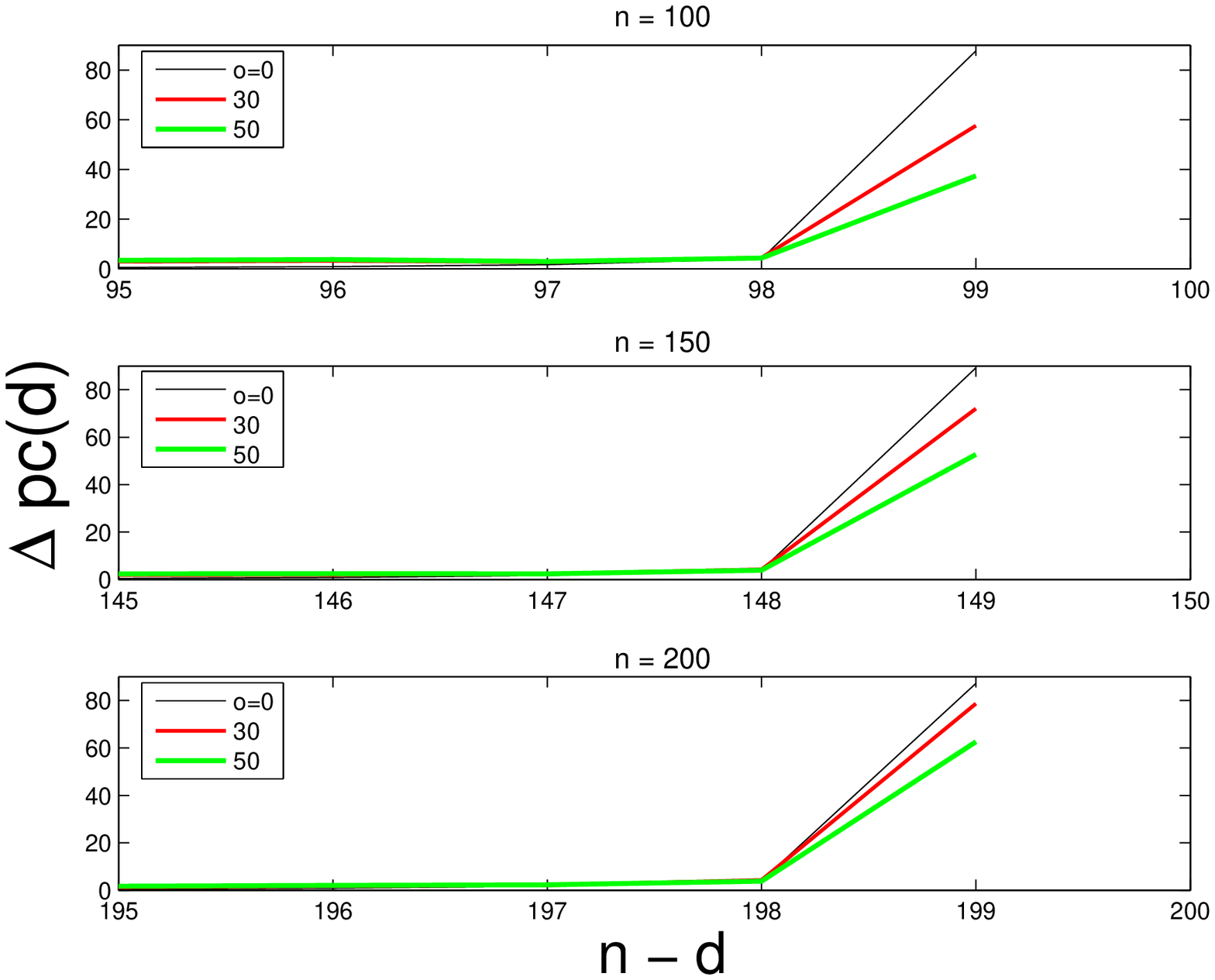}
\caption{\small Variations of $\D pc (d)$ with $n - t$ in the cases of GS ($o = 0$) and
ES (with $o = 30, 50$), and once again for $N = 300, n = 100, 150, 200$. The solid thin 
curve is for GS ($o = 0$), whereas the bold light and thick curves are respectively for 
ES with $o = 30$ and ES with $o = 50$.}
\la{fig:DOF-delta}
\end{center}
\vspace*{-0.55cm}
\end{figure*}
For GS, the entropy increases from $0$ to about $85 \%$ of the total entropy when the 
first lattice point just inside the horizon is included in the interaction region.
Inclusion of a second lattice point adds another $9 \%$, a third lattice point adds
$3 \%$, a fourth lattice point adds $1 \%$, and so on. The contributions to the entropy
by these additional lattice points farther and farther from the horizon decrease rapidly,
and by the time when the $(n/3)^{\rm th}$ is included the increment in entropy is less
than $0.01 \%$. This happens for all values of $n$, i.e., the horizon location does
not affect the way by which the DOF contribute to the entropy. For ES, however, the
inclusion of the first lattice point inside the horizon raises the entropy from $0$
to about $55 - 75 \%$ (depending on $n = 100 - 200$) for $o = 30$, and to about
$40 - 60 \%$ (depending on $n = 100 - 200$) for $o = 50$. The next successive points 
add about $9 \%, 4 - 5 \%, 3 - 4 \%, \cdots$, depending on $o = 30 - 50$ but fairly 
independent of $n$, and the corresponding slope is smaller.

From the above results we thus observe that although most of the entropy is contributed
by the DOF close to the horizon, the DOF that are farther away must also be taken into
account for the AL to emerge for GS, and the AL plus corrections for ES. With increasing 
excitations $o$, the contributions from the far-away DOF become more and more significant, 
as are the corrections to the AL. Thus the AL may be looked upon as a consequence of 
entanglement of the DOF near the horizon, whereas the corrections to the AL may be 
attributed to the contributions to the entropy by the DOF that are far from the horizon
\cite{sdshankiDoF,sdshankiss}. 

\section{Entanglement entropy of massive scalar field}  
\la{sf-mass}

In all the analysis presented in this review, we have considered an ideal situation
where the scalar field are non-interacting and massless. One natural question that 
arises is what happens if we include interactions or if the field is massive?
It is, in general, not possible to obtain $\Sen$ non-perturbatively for interacting 
fields and is beyond the scope of this review. In this section, we consider a 
massive field and show that all the results of the previous sections continue 
to hold.

The action for a massive scalar field propagating in the background space-time 
with metric $g_{\mu\nu}$ is given by
\be
S = -\frac{1}{2}
\int d^4 x \, \sqrt{-g}~ \le[g^{\mu\nu}~ \pa_{\mu}\vph~\pa_{\nu} \vph 
+ m^2 \vph^2 \ri]
\la{act-mass}
\ee
%

%
\begin{figure*}[!htb]
\begin{center}
\epsfxsize 4.75 in
\epsfysize 3.5 in
\epsfbox{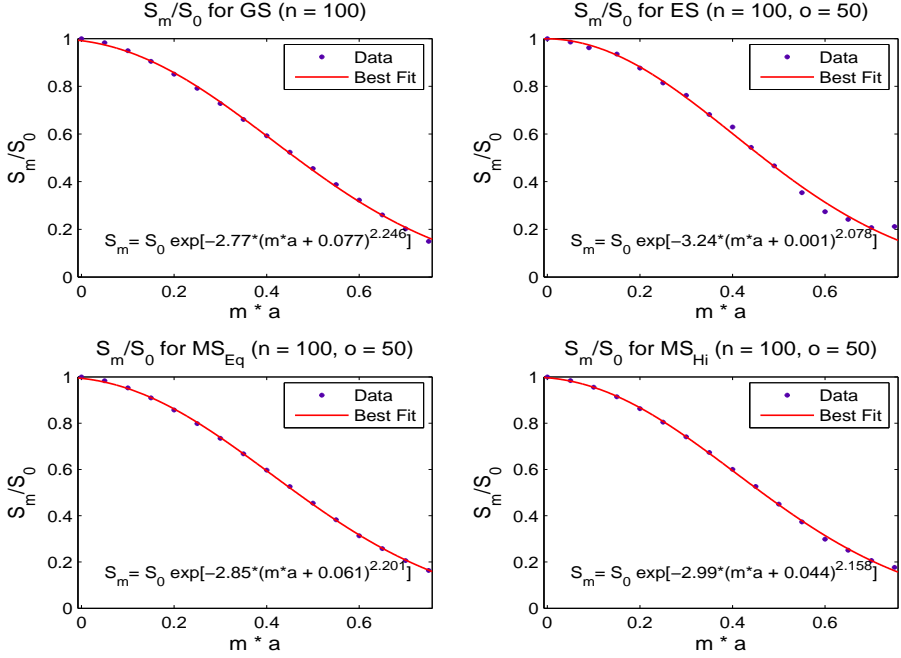}
\caption{\small Best fit plots of the relative variation of the total entropy 
$S_m$ for a massive scalar field (in units of the total entropy $S_0$
corresponding to a massless scalar field) with the mass $m$ times the 
lattice spacing $a$, for fixed $n = 100, o = 30$, in each of cases of 
GS, ES and MS (Eq/Hi). The corresponding data are shown by asterisks. 
The fits show an exponential damping of the ratio $S_m/S_0$ with mass.}
\la{fig:MS-mass}
\end{center}
\vspace*{-0.55cm}
\end{figure*}

Now, proceeding as before in the case of massless scalar field (see sec.
\ref{ent-sf}), one can obtain the discretized Hamiltonian for the case of
massive scalar field and show that this Hamiltonian resembles that of a
$N-$coupled HO, Eq. (\ref{coupledham1}), with an interaction matrix $K^{(m)}$
whose elements $K_{ij}^{(m)}$ are related to the interaction matrix elements 
$K_{ij}$ [Eq.(\ref{kij})] for a massless scalar field as 
\bea \la{kij-mass}
K_{ij}^{(m)} ~=~ K_{ij} ~+~ m^2 a^2 \, .
\eea
With this new interaction matrix $K^{(m)}$, one then finds the reduced density 
matrix following the steps discussed earlier and finally computes the entanglement 
entropy $S_{m}$ for the massive scalar field.

The variation of $S_m/S_0$ [where, $S_0$ is the entropy due to the massless scalar]
with $(m \times a)$ is shown in Fig. \ref{fig:MS-mass} for the cases of GS, ES and 
MS(Eq/Hi) for fixed parametric values $N = 300, n = 100, o = 50$. The data fit very
well with a Gaussian which shows that $S_m$ falls off exponentially with respect to 
$S_0$ as the mass increases:
\be
S_m ~=~ S_0 ~\exp\le[- \a_1 \le(m a ~+~ \a_2\ri)^\l\ri]
\ee
where $\a_1, \a_2$ and $\l$ are the fitting parameters. Depending on the state (GS, 
ES or MS), the parameter $\a_1$ varies between $2.77$ and $3.24$, $\a_2$ is between 
$0.077$ and $0.001$ and the power $\l$ is between $2.246$ and $2.078$. Therefore,
although $S_m/S_0$ approximately scales as $e^{- m^2 a^2}$, from the small variation
in the power $\l$ one finds that the exponential damping is strongest for GS, and 
gradually slows down as more and more ES oscillators are mixed with GS, the damping 
is slowest for the ES case.

Even with a fairly high amount of excitation ($o = 50$) the fitting parameters $\a_1, 
\a_2$ and $\l$ change very little for the different cases GS, MS(Eq/Hi) and ES. Therefore
for a particular value of the mass $m$, the relationship between $S_m$ and $(R/a)$
practically remains the same as that between $S_0$ and $(R/a)$ in all the cases of
GS, MS(Eq/Hi) and ES. The analysis and inferences of the previous sections go through 
for the massive scalar field as well, resulting in correction terms as obtained before. 

\section{Conclusions}       
\la{concl}

In the absence of a workable quantum theory of gravity, the best strategy is to slowly 
build a coherent picture and hope to understand --- and, in due course, solve --- 
some of the problems of black-hole thermodynamics. Thus, it is important to explore all 
possible avenues. Quantum entanglement,
as a source of black-hole entropy stands out for its simplicity and generality. The results 
discussed in this review highlight the nontrivial, and somewhat counterintuitive, 
facets of quantum entanglement and its role as the source of black-hole entropy. 
More precisely, assuming the modes evolve adiabatically, we have shown that:
\begin{itemize}
\item Entanglement leads to generic power-law corrections to the area law 
\item The quantum degrees of freedom that lead to $S_{_{\rm BH}}$ and
subleading corrections are different. 
\item It is possible to identify the quantum degrees of freedom that contribute 
to the area law and the subleading corrections. 
\item The interactions do not change the form of the entanglement entropy for 
different quantum states.
\end{itemize}

It is important to note that although the analysis presented here is semiclassical, 
since, the entanglement is a quantum effect and should be present in any theory 
of quantum gravity. Hence, the results presented here do have implications beyond 
the semiclassical regime. There are some new insights which arise in this approach 
which are worth exploring further:

\begin{itemize}
\item Is there a connection between the entanglement entropy and the Noether
charge? On the face of it there is no apparent connection: For  
diffeomorfism invariant theories, like Einstenian gravity, the Noether charge is 
interpreted as the black-hole entropy \cite{Wald:1993a}, implying that the 
higher order space-time derivatives contribute to the subleading power-law 
corrections. In the case of entanglement, the subleading contributions 
arise due to the excited quantum states of the scalar field which exists 
also for black-holes in Einstein gravity. 

However, it has been shown \cite{Frolov:1998vs} that the classical conserved charge for a 
nonminimally coupled matter fields propagating on the fixed 
curved background is identical to the Noether charge 
defined in Ref. \cite{Wald:1993a}. This raises the following question:
Can the excited states of the quantum scalar field be mapped to the 
non-minimal coupling of the field? If yes, then conserved charge
defined in Ref. \cite{Frolov:1998vs} can be related to the Noether 
charge. This is currently under investigation \cite{sdshankiss1}.

\item Any approach that aspires to explain black hole entropy from 
fundamental principles must provide a natural explaination for 
the factor $1/4$ in the Bekenstein-Hawking entropy. However, in the case 
of entanglement entropy, the proportionality constant 
in the relation $S = 0.3 (R/a)^2$ for GS
obtained in ref.\cite{sred} differs from the $1/4$ in the
Bekenstein-Hawking relation [Eq.(\ref{BH-law})].  This discrepancy
persists for MS and ES.  A probable reason behind this mismatch is the
dependence of the pre-factor on the type of the discretization scheme.
For example, another discretization scheme, resulting in the NN
interactions between four or more immediate neighbors, would result
in a different pre-factor. Is it at all possible to obtain the
Bekenstein-Hawking value? 

\item In this review, we have assumed that the quantum modes 
evolve adiabatically thus neglecting the contribution of the 
late-time modes leading to the Hawking particles. What happens if 
we relax this assumption? Will it change the relation (\ref{MS-fit})? 

The analysis in the previous section suggests that it may not, 
if we treat the late-time modes perturbatively. In Sec. (\ref{ent-sf}), 
we showed explicitly that the time-dependent Hamiltonian (\ref{eq:ham1})
becomes a free field Hamiltonian (\ref{sf-ham2}) for any Lem\^aitre 
time $\tau_0$. At any time $\tau = \tau_0 + \delta \tau$, the 
Hamiltonian (\ref{eq:ham1}) can be written as 
\be
H_{_{BH}} \simeq H_{_F} + \delta H 
\ee
where $\delta H$ is the small perturbation and includes the interaction. 
In this case, if  $(\delta H)/H_{_F} \ll 1$, the relation (\ref{MS-fit}) will 
not change. However, it will be interesting to investigate the effects 
at late times where the above perturbation expansion fails.

\item Recently, Pleino et al \cite{eisert} provided analytical proofs of 
numerical results of Bombelli et al \cite{bkls} and Srednicki \cite{sred} 
and showed that the entropy-area relation do not depend on the shape of the 
traced out volume \cite{eisert}. It will be interesting to do such an analysis 
for excited states. 

\item Can a temperature emerge in the entanglement entropy scenario, and 
if so, then along with the current entropy, will it be consistent with the 
first law of black hole thermodynamics?  Are the second and third laws of thermodynamics
valid for this entropy? Can the entanglement of scalar fields
help us to understand the evolution or dynamics of black-holes and the
information loss problem?  We hope to report on these in future.
\end{itemize}

\bigskip
\noindent
{\bf Acknowledgments:} 
The works of SD and SSu are supported by the Natural Sciences and 
Engineering Research Council of Canada. SSh is being 
supported by the Marie Curie Incoming International
fellowship IIF-2006-039205.


\begin{appendix}

\section{Appendix: Why consider scalar fields?}
\label{sf-mot}

In this appendix, we discuss the motivation for considering massless/massive scalar 
fields for the entanglement entropy computations, from the perspective of gravitational 
metric perturbations in asymptotically flat spherically symmetric space-times.

Let us consider the Einstein-Hilbert action with a positive cosmological constant 
$|\Lambda|$:
\be
\la{eq:EHAction}
S_{_{EH}} (\bar{g}) = M_{_{\rm Pl}}^2 \int d^4x \sqrt{-\bar{g}}
\le[\bar{R} - 2 |\Lambda|\ri] \, .  \ee
Let us decompose the metric $\bar{g}_{\mu\nu}$ in terms of a background metric
$g_{\mu\nu}$ and fluctuations $h_{\mu\nu}$:
\be
\bar{g}_{\mu\nu} = g_{\mu\nu} + h_{\mu\nu} \,.
\ee
Assuming $h_{\mu\nu}$ to be small, and expanding the action keeping only the 
parts quadratic in $h_{\mu\nu}$, gives \cite{barthchristen}
\bea
\la{eq:PerEH}
S_{_{EH}} (g, h) = - M_{_{\rm Pl}}^2 \! \int \!\! d^4x \sqrt{|g|}\, 
\le[ 2\, \gamma^\a_{~\mu\nu}\, \gamma_{~~\a}^{\mu\nu}  + 
\frac{1}{4}\nabla_{\mu}\tilde{h} \nabla^{\mu}\tilde{h} + 
\frac{|\Lambda|}{2} h_{\mu\nu} \tilde{h}^{\mu\nu} \ri] \, .
\eea
where
\bea
\la{htilde}
\tilde{h}_{\mu\nu} &\equiv& h_{\mu\nu}-\frac{1}{2} g_{\mu\nu}h_{\a}^{\a} \, ,
\qquad \tilde{h} \equiv \tilde{h}^{\mu}_{\mu} \, ,\\
\gamma^\a_{~\mu\nu} &\equiv& \frac{1}{2}
(\nabla_{\mu}\tilde{h}_{\nu}^{\a}+\nabla_{\nu}\tilde{h}_{\mu}^{\a} 
- \nabla^{\a}\tilde{h}_{\mu\nu}) \, .
\eea
One can easily verify that the above action (\ref{eq:PerEH}) is invariant under 
the infinitesimal gauge transformation $h_{\mu\nu} \to h_{\mu\nu} + \nabla_{(\mu} 
\xi_{\nu)}$ when the background metric $g_{\mu\nu}$ satisfies the vacuum Einstein's 
equation in presence of the cosmological constant $|\Lambda|$. The gauge arbitrariness 
can be removed by imposing the harmonic gauge condition $\pa_{\mu} \tilde{h}^{\mu\nu} 
= 0$ \cite{barthchristen}.

Now by keeping only the first derivatives of $h_{\mu\nu}$, the action (\ref{eq:PerEH}) 
further reduces to \cite{lem}:
\be \la{eq:PerEH1}
S_{_{EH}}(g, h) = - \frac{M_{_{\rm Pl}}^2}{2} \!\! \int \!\!\! d^4x \sqrt{|g|}\, 
\le[\nabla_{\alpha} {h}_{\mu\nu} \nabla^{\alpha} {h}^{\mu\nu} 
+ |\Lambda| h_{\mu\nu} h^{\mu\nu}\ri]\!\! \, .
\ee
This corresponds to the action for a massive spin-2 field $h_{\mu\nu}$ propagating in 
the background $g_{\mu\nu}$, the mass being given in terms of the cosmological constant 
$|\Lambda|$.

In the weak field limit ${h}^{\mu\nu}$ can be approximated as a plane-wave perturbation 
with a particular frequency, i.e., 
\be
h_{\mu\nu} = M_{_{\rm Pl}}^{-1} \epsilon_{\mu\nu} \varphi(x^{\mu}) \, ,
\ee
where $\epsilon_{\mu\nu}$ is the constant polarization tensor. Consequently, the above 
action (\ref{eq:PerEH1}) reduces to a form which is the same as the action for a massive 
scalar field $\vph$ propagating in the background $g_{\mu\nu}$:
\be
S_{_{EH}} (g, h) = - \frac 1 2 \int \!\!\! d^4x \sqrt{|g|}\, 
\le[\pa_{\alpha} \vph \pa^{\alpha} \vph 
+ |\Lambda| \vph^2 \ri] \, .
\ee

Now, one may further note that in four-dimensional spherically symmetric space-times,
the metric perturbations are of two kinds --- axial and polar \cite{chandra,qnm,perturb}. 
The equations of motion of both these perturbations are scalar in nature and are related 
to each other by a unitary transformation \cite{chandra}. The equations of motion of the 
axial perturbations are identical with those of a test, massless scalar field propagating 
in the black-hole background:
\be
\Box \vph \equiv \fr 1 {\sq{-g}} \pa_\m 
\le(\sq{- {g}} {g}^{\m\n} \pa_\n 
\vph\ri) = 0 \, .
\ee

Hence, by computing the entanglement entropy of the scalar fields one can obtain the 
entropy of a class of metric perturbations of the background space-time. Of course, such 
a computation would not account for the entropy of all perturbations, because a generic 
perturbation is a superposition of the plane wave modes and the entanglement entropy is 
a non-linear function of the wave-function. Nevertheless, scalar fields are expected to 
shed important light on the role of entanglement in the AL.

\section{Appendix: Hamiltonian of scalar fields in black-hole space-times}
\la{BH-Ham}

In this appendix, we find the expression for the Hamiltonian of a scalar field 
propagating in a static spherically symmetric space-time and show that for a particular 
time slicing this Hamiltonian reduces to that of a scalar field in flat space-time. 

Let us consider the line-element for a general four-dimensional spherically symmetric 
space-time:
\bea \la{bh-metric}
ds^2 &=& - A(\tau,\xi) \, d\tau^2 + \fr{d\xi^2}{B(\tau,\xi)} + 
\r^2(\tau,\xi) \le(d\theta^2 + \sin^2 \theta d\phi^2\ri) \, , 
\eea   
where $A, B, \r$ are continuous, differentiable functions of $(\tau,\xi)$. The action 
for the scalar field $\vph$ propagating in this space-time is given by
\bea \la{eq:actgen1}
S &=& -\fr 1 2 \int d^4 x \, \sqrt{-g}~ g^{\mu\nu}~ \pa_{\mu}\vph~\pa_{\nu}\vph \nn \\
&=& - \frac{1}{2} \sum_{l m} \int d\tau d\xi \Bigl[ -\frac{\rho^{2}}{\sqrt{A \, B}} 
(\pa_{\tau}\vph_{_{lm}})^2 + \sqrt{A B} \rho^{2} ~(\pa_{\xi}\vph_{_{lm}})^2 
+~ l(l + 1) \sqrt{\frac{A}{B}} \, \vph_{_{lm}}^2 \Bigr] \, . 
\eea
where we have decomposed $\vph$ in terms of the real spherical harmonics ($Z_{lm}
(\th, \f)$):
\be \la{tens-sph}
\vph (x^{\mu}) = \sum_{l m} \vph_{_{lm}}(\tau,\xi) Z_{l m} (\th, \f) \, . 
\ee
Following the standard rules, the canonical momenta and Hamiltonian of
the field are given by
\bea
\la{eq:mom}
{\Pi}_{_{lm}}&=& \frac{\pa \cal{L}}{\pa(\pa_{\tau} \vph_{lm})} =
\frac{\rho^{2}}{\sqrt{A \, B}} \, \pa_{\tau} \vph_{_{lm}} \, ,\\
\la{eq:gen-Ham}
H_{lm}(\tau) &=& \!\! \frac{1}{2} \int_{\tau}^{\infty} \!\!\!\!\!\!
d\xi \! \le[\! \frac{\sqrt{A B}}{\rho^{2}} \Pi_{_{lm}}^2 
+ \sqrt{A B} \, \rho^{2} (\pa_{\xi} \vph_{_{lm}})^2 + l(l + 1) \sqrt{\frac{A}{B}} \, 
\vph_{_{lm}}^2 \ri] \, , \qquad H = \sum_{lm} H_{lm} \, . 
\eea
The canonical variables $(\vph_{_{lm}}, \Pi_{_{lm}})$ satisfy the
Poisson brackets
\bea
\la{eq:gen-PB}
\{\vph_{_{lm}}(\tau,\xi), \Pi_{_{lm}}(\tau,\xi')\} = \delta(\xi - \xi') \quad ,
\qquad \{\vph_{_{lm}}(\tau,\xi), \vph_{_{lm}}(\tau,\xi')\} = 0 = 
\{\Pi_{_{lm}}(\tau,\xi), \Pi_{_{lm}}(\tau,\xi')\} \, .
\eea

In the time-dependent Lema\^itre coordinates \cite{lem,shanki:2k3} the metric components
of the line-element (\ref{bh-metric}) are given by
\be \la{eq:lemcoor}
A(\tau,\xi) = 1 \quad; \qquad B(\tau,\xi) = \frac{1}{1 - f(r)} \quad; \qquad
\rho(\tau,\xi) = r \, ,
\ee
where $r = r(\tau,\xi)$. The line-element in the Lema\^itre coordinates is related to 
that in the time-independent Schwarzschild coordinates, viz.,
\be \la{sch1}
ds^2 = - f(r) dt^2 + \fr{dr}{f(r)} + r^2 \le(d\theta^2 + \sin^2 \theta d\phi^2\ri) \quad;  
\qquad  f(r = r_h) = 0
\ee
by the following transformations \cite{shanki:2k3}:
\bea \la{eq:xitau}
\tau = t \pm \int\!\! dr \frac{\sqrt{1 - f(r)}}{f(r)} \quad; \qquad 
\xi = t + \int\!\! dr \frac{[1 - f(r)]^{-1/2}}{f(r)} \quad, \qquad
\xi - \tau &=& \int \frac{dr}{\sqrt{1 - f(r)}} \, .
\eea
Unlike the line-element in Schwarzschild coordinates, the line-element in Lema\^itre 
coordinates is not singular at the horizon $r_h$. Moreover, the coordinate  $\xi$
(or, $\tau$) is space(or, time)-like everywhere, whereas $r$(or, $t$) is space(or, 
time)-like only for $r > r_h$.  

In the Lema\^itre coordinates the general Hamiltonian (\ref{eq:gen-Ham}) takes the
form
\bea
\la{eq:lem-Ham}
H_{_{lm}}(\tau) ~=~ \frac{1}{2} \int_{\tau}^{\infty} d\xi \le[ \frac{1}{r^2 
\sqrt{1 - f(r)}} \Pi_{_{lm}}^2 + \frac{r^{2}}{\sqrt{1 - f(r)}} \le(\pa_{\xi} 
\vph_{_{lm}}\ri)^2  + l(l + 1)\sqrt{1 - f(r)} \, \vph_{_{lm}}^2 \ri] \, ,
\eea
which depends explicitly on the Lema\^itre time. 

Choosing now a fixed Lema\^itre time ($\tau = \tau_0 = 0$, say), the relations 
(\ref{eq:xitau}) lead to:
\be \la{dxi}
\frac{d\xi}{dr} = \frac{1}{\sqrt{1 - f(r)}} \, .
\ee
If we set $d\theta = d\phi = 0$, then for the fixed Lema\^itre time $\tau_0$
it follows that $ds^2 = d\xi^2/B(\tau_0,\xi) = dr^2$, i.e., the covariant cut-off 
is $|ds| = dr$. Substituting the above relation (\ref{dxi}) in the Hamiltonian
(\ref{eq:lem-Ham}) we get
\be
H_{_{lm}}(0) = \fr 1 2 \int_{0}^{\infty} dr \le[\frac{\Pi_{_{lm}}^2 r^{-2}}{1 - f(r)} 
+ r^2 \le(\pa_r \vph_{_{lm}}\ri)^2 + l(l + 1) \, \vph_{_{lm}}^2\ri] \, ,
\ee
where the variables $(\vph_{_{lm}}, \Pi_{_{lm}})$ satisfy the relation:
\be
\{\vph_{_{lm}}(r), \Pi_{_{lm}}(r')\} = \sqrt{1 - f(r)} \delta(r - r') .
\ee

Performing the following canonical transformations
\be
\Pi_{_{lm}} \to {r \sqrt{1 - f(r)}} \, \Pi_{_{lm}}  \, ; \,  
\vph_{_{lm}} \to \frac{\vph_{_{lm}}}{r} 
\ee
the full Hamiltonian reduces to that of a free scalar field propagating
in flat space-time \cite{melnikov}
\bea
H = \sum_{lm} \fr 1 2 \int_0^\infty dr 
\le\{\p_{lm}^2(r) + r^2 \le[\fr{\pa}{\pa r} \le(\fr{\varphi_{lm} 
(r)}{r}\ri)\ri]^2 + \fr{l(l+1)}{r^2}~\varphi_{lm}^2(r)\ri\} \, . 
\la{ham2}
\eea
This happens for {\it any} fixed value of the Lema\^itre time $\t$, provided the 
scalar field is traced over either the region $r \in (0, r_h]$ or the region $r 
\in [r_h, \infty)$. Note that the black-hole singularity can be entirely avoided for 
the latter choice, and for evaluating time-independent quantities such as entropy, it
suffices to use the above Hamiltonian.

The approach here differs from that of Ref. \cite{mukoh} where the
authors divide the exterior region $r \geq r_s$ into two by
introducing an hypothetical spherical surface and obtain the
entanglement entropy of that surface. In contrast, we consider the
complete $r \geq r_s$ region and obtain the entropy for the black hole horizon.

\end{appendix}

\end{document}